\documentclass[aps, pra, onecolumn, tightenlines, letterpaper, amsmath, amssymb, preprintnumbers, floatfix, longbibliography, nofootinbib]{revtex4-2}

\usepackage{dsfont}
\usepackage{amsmath}
\usepackage{amssymb}
\usepackage{physics}
\usepackage{tabu}
\usepackage{tabularx}
\usepackage{bm}
\usepackage{tikz}
\usetikzlibrary{quantikz}
\usepackage{graphicx}
\usepackage{placeins}
\usepackage{textgreek}
\usepackage{multirow}
\usepackage{makecell}
\usepackage{soul}
\usepackage{diagbox}

\usepackage[normalem]{ulem}

\makeatletter
\newcommand{\fmarki}{*}
\newcommand{\fmarkii}{\ensuremath{\dagger}}
\newcommand{\fmarkiii}{\ensuremath{\ddagger}}
\newcommand{\fmarkiv}{\ensuremath{\mathsection}}
\newcommand{\fmarkv}{\ensuremath{\mathparagraph}}
\newcommand{\fmarkvi}{\ensuremath{\|}}

\newcommand{\infiL}{{\mathcal{I}_L}}

\def\@fnsymbol#1{{\ifcase#1\or \fmarki\or \fmarkii\or \fmarkiii\or \fmarkiv\or \fmarkv\or \fmarkvi \else\@ctrerr\fi}}
\makeatother

\renewcommand{\fmarkvi}{\$}

\usepackage[hypertexnames=false]{hyperref}
\hypersetup{
    colorlinks=true,       
    linkcolor=blue,          
    citecolor=blue,        
    filecolor=blue,      
    urlcolor=blue           
}

\def\Dslash{{ D\hskip-0.6em /}}

\newcolumntype{Y}{>{\centering\arraybackslash}X}

\makeatletter
\pretocmd\frontmatter@thefootnote{\color{black}}{}{}
\makeatother

\usepackage{orcidlink}

\begin{document}

\begin{figure}
  \vskip -1.cm
  \leftline{\includegraphics[width=0.15\textwidth]{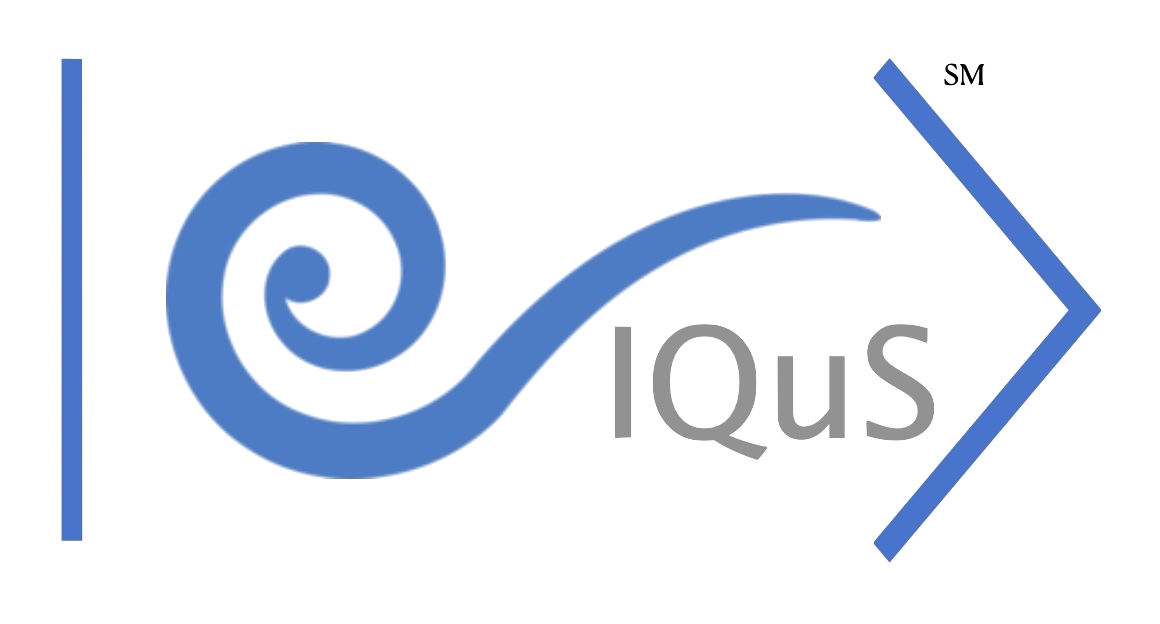}}
\end{figure}

\title{Scalable Circuits for Preparing Ground States on Digital Quantum Computers: \\
The Schwinger Model Vacuum on 100 Qubits}

\author{Roland C.~Farrell\,\orcidlink{0000-0001-7189-0424
}}
\email{rolanf2@uw.edu}
\affiliation{InQubator for Quantum Simulation (IQuS), Department of Physics, University of Washington, Seattle, WA 98195, USA.}
\author{Marc Illa\,\orcidlink{0000-0003-3570-2849}}
\email{marcilla@uw.edu}
\affiliation{InQubator for Quantum Simulation (IQuS), Department of Physics, University of Washington, Seattle, WA 98195, USA.}
\author{Anthony N. Ciavarella\,\orcidlink{0000-0003-3918-4110}}
\email{aciavare@uw.edu}
\affiliation{InQubator for Quantum Simulation (IQuS), Department of Physics, University of Washington, Seattle, WA 98195, USA.}
\author{Martin J.~Savage\,\orcidlink{0000-0001-6502-7106}}
\email{mjs5@uw.edu}
\affiliation{InQubator for Quantum Simulation (IQuS), Department of Physics, University of Washington, Seattle, WA 98195, USA.}

\preprint{IQuS@UW-21-060, NT@UW-23-13}
\date{\today}

\begin{abstract}
\noindent
The vacuum of the lattice Schwinger model is prepared on up to 100 qubits of IBM's {\tt Eagle}-processor quantum computers.
A new algorithm to prepare the ground state of a gapped translationally-invariant system
on a quantum computer  is presented,
which we call 
Scalable Circuits ADAPT-VQE (SC-ADAPT-VQE).
This algorithm uses the exponential decay of correlations between distant regions of the ground state, together with ADAPT-VQE, 
to construct quantum circuits for state preparation that can be scaled to arbitrarily large systems.
These scalable circuits can be determined using classical computers, avoiding the challenging task of optimizing parameterized circuits on a quantum computer.
SC-ADAPT-VQE is applied to the Schwinger model, 
and shown to be systematically improvable, with an accuracy that converges exponentially with circuit depth.
Both the structure of the circuits and the deviations of prepared wavefunctions are found to become independent of the number of spatial sites, $L$.
This allows for a controlled extrapolation of the circuits, 
determined using small or modest-sized systems,
to arbitrarily large $L$.
The circuits 
for the Schwinger model are determined on lattices up to $L=14$ (28 qubits) with 
the {\tt qiskit} classical simulator, 
and subsequently scaled up to prepare the $L=50$ (100 qubits) vacuum on IBM's 127 superconducting-qubit quantum computers 
{\tt ibm\_brisbane} and {\tt ibm\_cusco}.
After introducing an improved error-mitigation technique, 
which we call Operator Decoherence Renormalization,
the chiral condensate and charge-charge correlators obtained from 
the quantum computers are found to be in good agreement with classical Matrix Product State simulations.
\end{abstract}

\maketitle
\newpage{}

\begingroup
\hypersetup{linkcolor=black}
\tableofcontents
\endgroup

\pagenumbering{gobble}

\newpage{}
\section{Introduction}
\pagenumbering{arabic}
\setcounter{page}{1}
\label{sec:I}
\noindent
Quantum simulations of physical systems described by the Standard Model~\cite{Weinberg:1967tq,Glashow:1961tr,Salam:1968rm,Politzer:1973fx,Gross:1973id,Higgs:1964pj}, 
and descendant effective field theories (EFT),
are anticipated to provide qualitatively new predictions about matter under extreme conditions; from the dynamics of matter in the early universe, 
to properties of the exotic phases of quantum chromodynamics (QCD) produced at the LHC and RHIC (for overviews and reviews, see Refs.~\cite{Banuls:2019bmf,Guan:2020bdl,Klco:2021lap,Delgado:2022tpc,Bauer:2022hpo,Humble:2022klb,Beck:2023xhh,Bauer:2023qgm,DiMeglio:2023nsa}).
One of the major challenges facing quantum simulations of physical systems is 
the preparation of initial states on quantum computers
that can be used to determine important quantities that are inaccessible to  
classical high-performance computing (HPC),
i.e., the problem of state preparation.
While simulating the dynamics of any given initial state 
is known to be efficient for an ideal quantum computer~\cite{Lloyd1073},
residing in the {\bf BQP} complexity class,
preparing an arbitrary state generally requires 
quantum resources that asymptotically scale 
super-polynomially  with increasing system size~\cite{10.1007/978-3-540-30538-5_31},
residing in the {\bf QMA} complexity class.\footnote{Note that adiabatic state preparation resides within {\bf BQP} when there is a path through parameter space in which the system remains gapped~\cite{farhi2000quantum,van_Dam_2001}.
However, even in gapped systems the gate count required for adiabatic preparation can be daunting; e.g., see Ref.~\cite{Chakraborty:2020uhf} where adiabatic preparation of the Schwinger model vacuum on 16 qubits was estimated to require $2.7\times10^5$ two-qubit gates.}
However, states of physical systems are not the general case, and are often constrained by both local and global symmetries.\footnote{Systems of importance to nuclear physics and high-energy physics are constrained by
a number of local, exact global and approximately global symmetries, 
some of which are emergent from the mechanisms of confinement 
and spontaneous symmetry breaking.} 
In some instances, these symmetries
allow observables to be computed by perturbing around states that can be efficiently 
initialized~\cite{Klco:2021lap}.
In the foreseeable future, quantum simulations  will be far from asymptotic in both system size and evolution time, and the resources required for both time evolution and state preparation
will be estimated by direct construction and extrapolations thereof.
Furthermore, successful quantum simulations will require specialized quantum circuits and workflows that are optimized for specific quantum hardware.

The development of algorithms for preparing 
non-trivial initial states on quantum computers, including the ground states of quantum field theories (QFTs), is an active area of research.
Even with many advances, 
algorithms remain limited in capability, and generally do not scale favorably to modest or large-scale simulations of quantum many-body systems.
Consequently, quantum simulations of small model systems are currently 
being performed across an array of science domains,
generally studying dynamics starting from tensor-product initial states.
While being the simplest gauge theory based on a continuous group, 
the Schwinger model~\cite{Schwinger:1962tp} (quantum electrodynamics in 1+1D)  
possesses many features of interest to both the 
quantum chromodynamics and quantum information science (QIS)
communities.
These include
the presence of a mass gap, charge screening, a chiral condensate, few-body bound states (``hadrons'' and ``nuclei''), and  a topological $\theta$-term.
It has emerged as a popular test bed for developing quantum simulation techniques
for  lattice gauge theories,
and has been explored using a variety of platforms, including trapped ions~\cite{Martinez:2016yna,Kokail:2018eiw,Nguyen:2021hyk,Mueller:2022xbg}, superconducting qubits~\cite{Klco:2018kyo,Mazzola:2021hma,deJong:2021wsd,Gong:2021bcp,Mildenberger:2022jqr,Charles:2023zbl,Pomarico:2023png}, photonic systems~\cite{Lu:2018pjk}, Rydberg atoms~\cite{Surace:2019dtp}, ultracold atoms~\cite{Mil:2019pbt,Yang:2020yer,Zhou:2021kdl,PhysRevResearch.5.023010,zhang2023observation} and classical electric circuits~\cite{Riechert:2021ink}, together with classical simulations~\cite{Banerjee:2012pg,Marcos:2013aya,Yang:2016hjn,Muschik:2016tws,Davoudi:2019bhy,Luo:2019vmi,Chakraborty:2020uhf,Ferguson:2020qyf,Davoudi:2021ney,Yamamoto:2021vxp,Honda:2021aum,Bennewitz:2021jqi,Andrade:2021pil,Honda:2021ovk,Halimeh:2022pkw,Xie:2022jgj,Davoudi:2022uzo,Nagano:2023uaq,Popov:2023xft,Nagano:2023kge}, calculations~\cite{Hauke:2013jga,Kasper:2016mzj,Notarnicola:2019wzb,Tran:2020azk,Shen:2021zrg,Jensen:2022hyu,Florio:2023dke,Ikeda:2023zil} 
and tensor-networks~\cite{Byrnes:2002nv,Banuls:2013jaa,Rico:2013qya,Buyens:2013yza,Kuhn:2014rha,Banuls:2015sta,Pichler:2015yqa,Buyens:2015tea,Banuls:2016lkq,Buyens:2016ecr,Buyens:2016hhu,Zapp:2017fcr,Ercolessi:2017jbi,Sala:2018dui,Funcke:2019zna,Magnifico:2019kyj,Butt:2019uul,Rigobello:2021fxw,Okuda:2022hsq,Honda:2022edn,Desaules:2023ghs,Angelides:2023bme,Belyansky:2023rgh} (for reviews on this last topic, see, e.g., Refs.~\cite{Banuls:2019rao,Banuls:2019bmf}).
There has also been pioneering work on quantum simulations of low-dimensional 
non-Abelian gauge theories, both with~\cite{Atas:2021ext,Atas:2022dqm,Farrell:2022vyh,Farrell:2022wyt,Ciavarella:2023mfc,Kuhn:2015zqa} 
and without~\cite{Klco:2019evd,Ciavarella:2021nmj,Ciavarella:2021lel,ARahman:2021ktn,Illa:2022jqb,Ciavarella:2022zhe,ARahman:2022tkr} matter.
While these are important benchmarks, more sophisticated simulations requiring the preparation of eigenstates or scattering states have so far been too demanding for 
NISQ-era quantum computers, and until now have been limited to 20 qubits~\cite{Kokail:2018eiw,Mildenberger:2022jqr}.

Many systems of physical interest, including QCD, 
have translational symmetry and possess an energy (mass) gap $\Lambda$ between the unique ground state and first excited state.
The  gap defines a characteristic length scale of the system $\xi = 1/\Lambda$, and parameterizes the decay of the longest distance 
correlations in the ground state wavefunction, falling as 
$\sim e^{- r/\xi}/r^\alpha$ for regions separated by $r\gg\xi$, for some $\alpha$.
A natural way to encode a lattice QFT onto a register of a digital quantum computer is by identifying subsets of qubits (or qudits) with spatial points of the lattice 
that align with the connectivity of the quantum computer.
A realization of the ground state on the register of a quantum computer
should reflect the localized correlations
between these subsets of 
qubits
separated by $r\gg\xi$~\cite{Klco:2019yrb,Klco:2020aud}.
In the absence of topological order, one way to establish the ground state is to initialize the quantum register in a 
state without correlations between qubits, e.g., a tensor product state,
and 
then systematically introduce correlations 
through the action of  quantum circuits.
A crucial point is that the localized correlations imply that the state preparation circuits need to have structure only for qubits spatially separated by $r \lesssim \xi$~\cite{Klco:2019yrb,Klco:2020aud}.
This is sufficient to obtain exponentially converged accuracy in the prepared state.
Due to translational invariance, the ground state for an arbitrarily large lattice can be prepared by repeating these circuits across the entire register.

\begin{figure}[!bth]
    \centering
    \includegraphics[width=\columnwidth]{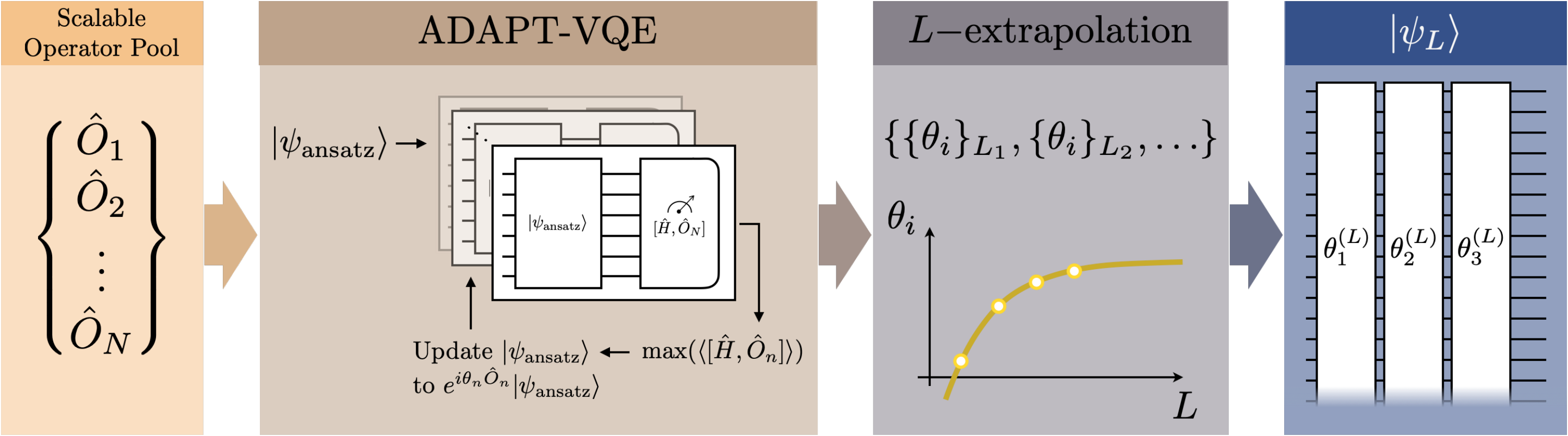}
    \caption{Pictorial description of the SC-ADAPT-VQE algorithm. Once a pool of scalable operators $\{\hat{O}_i\}$ has been identified, 
    ADAPT-VQE is performed using {\it classical} computers to determine a quantum circuit (parameterized by $\{ \theta_i \}$) that prepares the vacuum up to a desired tolerance.
    ADAPT-VQE is repeated for multiple lattice sizes, $\{L_1,L_2,...\}$, and the circuit parameters are extrapolated to the desired $L$, which can be arbitrarily large.
    The extrapolated circuits are executed on a quantum computer to prepare the vacuum on a system of size $L$.}
    \label{fig:SCADAPTVQE-diag}
\end{figure}

To study the dynamics of physically relevant systems in a quantitative way,
with a complete quantification of uncertainties,
simulations of large volumes of spacetime are typically required.
Motivated by the discussion in the previous paragraph, we introduce Scalable Circuits ADAPT-VQE (SC-ADAPT-VQE), a new method for quantum state preparation that uses the hierarchies of length scales present in physical systems; see Fig.~\ref{fig:SCADAPTVQE-diag} for an illustration.
In SC-ADAPT-VQE, 
quantum circuits that (efficiently) prepare a given state to a specified level of precision
are determined on modest-sized lattices that are large enough to contain the longest correlation lengths.
As long as $\xi$ is not too large, these circuits can be determined using {\it classical} computers.
This avoids the challenging task of optimizing circuits on a quantum computer with both statistical uncertainty and device noise~\cite{Wang:2020yjh,Scriva:2023sgz}.
Once determined, (discrete) translation invariance is used to scale these circuits up to the full lattice.
Since the quality of the prepared state becomes 
independent of the spatial lattice length $L$, 
with ${\mathcal O}(e^{-\xi/L})$ corrections,
this is a potential path toward 
quantum simulations of lattice QFTs
that are beyond the capabilities of HPC.

In this work, SC-ADAPT-VQE is applied to the Schwinger model and is used to prepare the vacuum on up to 100 qubits on IBM's {\tt Eagle} quantum processors.
Underlying the development is the ADAPT-VQE algorithm~\cite{Grimsley_2019} for quantum state preparation, which is modified to generate scalable circuits.
After the necessary Trotterized circuits have been built, SC-ADAPT-VQE is performed using the {\tt qiskit} classical simulator on 
system sizes up to $L=14$ (28 qubits).
It is found that both the energy density and the chiral condensate 
converge exponentially with circuit depth to the exact results.
Importantly, both the quality of the prepared state and the structure of the associated circuits are found to converge with system size.
This allows the state preparation circuits, determined on small lattices using classical computing, to be extrapolated to much larger lattices, with a quality that becomes independent of $L$.
The scaled circuits are used to prepare the $L\le 500$ vacua using {\tt qiskit}'s Matrix Product State (MPS) circuit simulator, 
and to prepare the  $L \le 50$ (100 qubits) vacua on the registers of 
IBM's superconducting-qubit quantum computers {\tt ibm\_brisbane} and {\tt ibm\_cusco}.
An improved and unbiased error mitigation technique, Operator Decoherence Renormalization (ODR), is developed and applied to the quantum simulations to estimate error-free observables.
The results obtained from both the MPS circuit simulator and  IBM's quantum computers
are found to be in excellent agreement with Density Matrix Renormalization Group (DMRG) calculations.
%

\section{The Lattice Schwinger Model}
\label{sec:SchwingerHam}
\noindent
The Schwinger model~\cite{Schwinger:1962tp} has a long history of study 
in the continuum and using numerical lattice techniques.
 In the continuum, it is described by the Lagrange density
 \begin{align}
 {\cal L}
 & =  \overline{\psi}\left( i \Dslash - m_\psi \right) \psi - \frac{1}{4} F^{\mu\nu} F_{\mu\nu}
 \ .
 \label{eq:LagSM}
 \end{align}
Electrically-charged fermions are described by the field operator $\psi$ with mass $m_\psi$,
the electromagnetic gauge field by $A_\mu$ with field tensor $F_{\mu\nu}$, 
and the covariant derivative is defined as $D_\mu = \partial_\mu - i e A_\mu$.
It is the Hamiltonian lattice formulation,
first developed and studied by Banks, Kogut and Susskind~\cite{Kogut:1974ag,Banks:1975gq},
that is relevant for quantum simulations. 
One feature of gauge theories in $1+1$D, which distinguishes 
them from theories in higher dimensions,
is that the gauge field is completely constrained 
by the distribution of fermion charges through Gauss's law.
In axial gauge, the spatial gauge field is absent, and the effects of the time-component of the gauge field
are included by non-local (Coulomb) interactions~\cite{Sala:2018dui,Farrell:2022wyt}.
With  open boundary conditions (OBCs),  
using the staggered fermion discretization~\cite{Kogut:1974ag} of the electron field,
and applying the Jordan Wigner (JW) transformation to map fermion field operators to spins, 
the Schwinger model Hamiltonian is (for a derivation, see, e.g., Ref.~\cite{Kokail:2018eiw})
\begin{align}
\hat H & \ =\  \hat H_m + \hat H_{kin} + \hat H_{el} \ = \ \frac{m}{ 2}\sum_{j=0}^{2L-1}\ \left[ (-1)^j \hat Z_j + \hat{I} \right] \ + \ \frac{1}{2}\sum_{j=0}^{2L-2}\ \left( \hat \sigma^+_j \hat\sigma^-_{j+1} + {\rm h.c.} \right) \ + \ \frac{g^2}{ 2}\sum_{j=0}^{2L-2}\bigg (\sum_{k\leq j} \hat Q_k \bigg )^2 
\ ,
\nonumber \\ 
\hat Q_k & \ = \ -\frac{1}{2}\left[ \hat Z_k + (-1)^k\hat{I} \right] \ .
\label{eq:Hgf}
\end{align}
Here, $L$ is the number of  spatial lattice sites, 
corresponding to $2L$ staggered (fermion) sites, $m$ and $g$ are the (bare) electron mass and charge, respectively, and the staggered lattice spacing $a$ has been set to one.\footnote{For faster convergence to the continuum, an ${\mathcal O}(a)$ improvement to the mass term can be performed to restore a discrete remnant of chiral symmetry in the $m\to0$ limit~\cite{Dempsey:2022nys}.} 
``Physical'' quantities are 
derived from the corresponding dimensionless quantities by restoring factors of the 
spatial lattice spacing.
Even (odd) sites correspond to electrons (positrons), as reflected in the staggered mass term and charge operator.\footnote{The convention is that even fermion-sites correspond to electrons, 
such that $\hat Q \lvert\downarrow\rangle = 0$ and 
$ \hat Q \lvert\uparrow\rangle = -\lvert\uparrow\rangle$,
while
the odd fermion-sites correspond to positrons, such that 
$ \hat Q \lvert\uparrow\rangle = 0$ and 
$\hat Q \lvert\downarrow\rangle = +\lvert\downarrow\rangle$.
}
A background electric field can be included straightforwardly, 
equivalent to a  $\theta$-term,  but
will be set to zero in this work.
In the sector with vanishing total electric charge, $\hat{H}_{el}$ 
in Eq.~(\ref{eq:Hgf})
can be rewritten 
in a way that reduces the number of gates required in quantum circuits for time evolution, and is less demanding on device connectivity, see App.~\ref{app:QoSM}.
Due to the confinement, the low energy excitations are hadrons and the mass gap is given by $\Lambda = m_{\text{hadron}}$. For our purposes, $m_{\text{hadron}}$ is defined to be the energy difference in the $Q=0$ sector between the first excited state (single hadron at rest) and the vacuum.

\subsection{Infinite Volume Extrapolations of Local Observables}
\label{sec:infVolExtra}
\noindent
Central to the development of state preparation circuits is the scaling of expectation values of local observables in the ground state, 
with both the correlation length $\xi = 1/m_{{\rm hadron}}$, and the volume $L$.
Due to the exponential suppression of correlations in the ground state 
between regions separated by $r>\xi$, it is expected that, locally, the wavefunction has converged to its infinite volume form, with corrections of ${\mathcal O}(e^{-\xi/L})$.
As a result, expectation values of local observables will 
be exponentially converged to their infinite volume values.
However, near the boundaries of the lattice, the wavefunction is perturbed over a depth proportional to $\xi$, causing local observables to deviate from their infinite volume values.
Equivalently, boundary effects cause deviations in volume averages of local observables that are ${\mathcal O}(\xi/L)$.
This scaling of observables is responsible for the SC-ADAPT-VQE prepared vacuum converging exponentially in circuit depth, and enables the circuits to be systematically extrapolated to larger system sizes.

\begin{figure}[hbt!]
    \centering
    \includegraphics[width=\columnwidth]{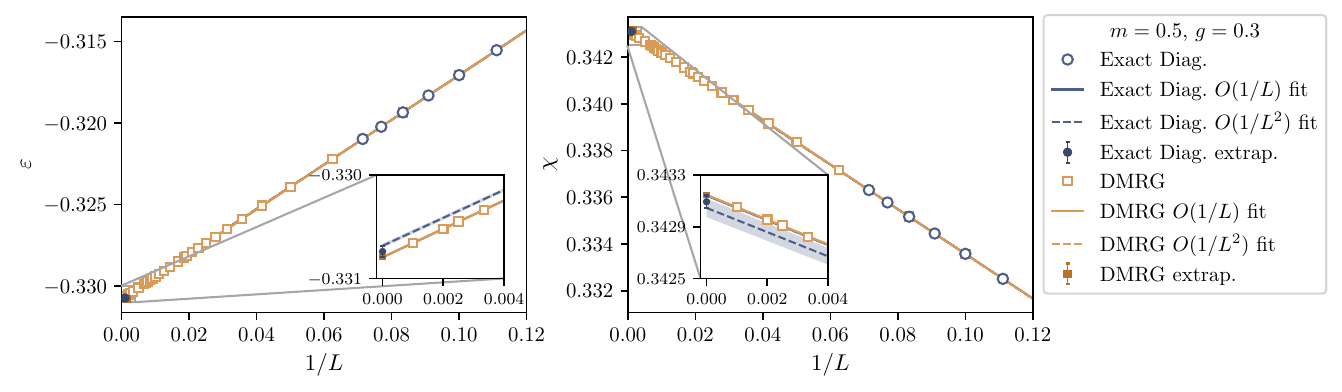}
    \caption{
    $L$-extrapolations of the vacuum energy density $\varepsilon$ (left panel)
and chiral condensate $\chi$ (right panel) for $m=0.5$ and $g=0.3$.
The results of exact diagonalization calculations for $L\ge 9$ (blue circles) given in 
Table~\ref{tab:Echifmp5gp3} and 
DMRG calculations (orange squares) given in Table~\ref{tab:DMRGresults}
are extrapolated to $L\rightarrow\infty$, as shown by the darker points, respectively.
The solid lines correspond to linear extrapolations and the dashed lines correspond to quadratic extrapolations, and are found to overlap (see inset).   
The difference between the $L\rightarrow\infty$ values of these two extrapolations defines the (fitting) uncertainties associated with the darker points. 
}
    \label{fig:EdChiLextrap53}
\end{figure}

Two quantities associated with the ground-state wavefunction (vacuum) 
that we focus on are the chiral condensate $\chi$, and the energy density
$\varepsilon$.
The chiral condensate\footnote{In the continuum, 
the chiral condensate is defined as 
$\chi_{\text{cont}}=\langle \overline{\psi}\psi \rangle$, which on the lattice becomes $\chi_{\text{lat}} = \frac{1}{L}\sum_j\langle \overline{\psi}_j\psi_j \rangle$, 
where $j$ labels the spatial site. 
To have a positive quantity, we have added a constant to the definition of $\chi$, 
$\chi \equiv \chi_{\text{lat}} + 1$. 
This counts the average number of electrons and positrons on a spatial site.} 
is an order parameter of chiral symmetry breaking, 
and in the JW mapping is
\begin{equation}
\chi \ = \ \frac{1}{2L}  \sum_{j=0}^{2L-1} \left\langle (-1)^j \hat Z_j + \hat I   \right\rangle \ \equiv \ \frac{1}{2L}  \sum_{j=0}^{2L-1}\chi_j
\ .
\label{eq:chiralCond}
\end{equation} 
The energy density is defined as 
$\varepsilon = \langle \hat{H} \rangle / L$, 
and 
in axial gauge is not a local observable because the contribution from the 
electric-field term in the Hamiltonian, $\hat{H}_{el}$, involves all-to-all couplings.
However, this is an artifact of using axial gauge and enforcing Gauss's law.
In Weyl gauge, with explicit (local) gauge degrees of freedom, 
the Hamiltonian is manifestly local, and therefore the energy density is a local observable.
These quantities are computed for $m=0.5,g=0.3$ using exact diagonalization for $L\le 14$ 
(Table~\ref{tab:Echifmp5gp3})
and DMRG for $L\gg 14$ (Table~\ref{tab:DMRGresults}).
As anticipated,
a linear extrapolation in $1/L$ is found to be consistent with 
these results, as seen in Fig.~\ref{fig:EdChiLextrap53}.
Additional details,
along with results for $m=0.1$  with $g=0.3$ and $g=0.8$, 
can be found in App.~\ref{app:LinvScaling}.

\FloatBarrier
\section{SC-ADAPT-VQE for the Lattice Schwinger Model}
\label{sec:II}
\noindent
Underlying SC-ADAPT-VQE is ADAPT-VQE~\cite{Grimsley_2019},  
a quantum algorithm for state preparation that has been applied to spin models~\cite{VanDyke:2022ffj}, systems in quantum chemistry~\cite{Grimsley_2019,Tang:2021PRX,Yordanov:2021,Shkolnikov:2021btx,Bertels:2022oga,Anastasiou:2022swg,Feniou:2023gvo} and nuclear structure~\cite{Romero:2022blx,Perez-Obiol:2023vod}.
It builds upon the Variational Quantum Eigensolver (VQE)~\cite{Peruzzo_2014}, 
in which parameterized quantum circuits are optimized  
to minimize the expectation value of a Hamiltonian.
The parameterized circuits are constructed 
step-wise (or equivalently in layers), 
where the incrementally-improved ansatz states converge to the ground state with successive iterations.
At each step, 
the unitary transformation that maximally decreases the energy of the ansatz state is identified from a pre-defined set (``pool") of unitaries.
The quantum circuit corresponding to this unitary is then
appended to the state preparation circuit. 
The (initial) state from which the algorithm starts
will often be chosen to be a tensor product
or an entangled state that can be efficiently prepared on a quantum computer, 
such as a GHZ-state.
If the target state is the ground state of a confining gauge theory, e.g., the Schwinger model, 
the strong-coupling (trivial) vacuum,
\begin{equation}
|\Omega_0\rangle = 
\lvert\uparrow\downarrow\uparrow\downarrow \ldots\uparrow\downarrow\rangle
\ ,
\label{eq:SCvac}
\end{equation} 
can be a good choice for such an initial state as it has the correct long-distance 
structure in the gauge fields.
The ADAPT-VQE algorithm can be summarized as follows:
\begin{itemize}
    \item[1.] Define a pool of operators $\{ \hat{O} \}$ that are 
    constrained to respect some or all of the symmetries of the system. 
    \item[2.] Initialize the register of the quantum computer to a strategically selected state, 
    $|\psi_{{\rm ansatz}} \rangle$,  with the desired quantum numbers
   and symmetries of the target wavefunction.
    \item[3.] Measure the expectation value of the commutator of the Hamiltonian with each operator in the pool, 
    $\langle \psi_{{\rm ansatz}} \lvert [\hat{H}, \hat{O}_i] \lvert \psi_{{\rm ansatz}} \rangle$. 
    These are estimators of the associated decrease in energy from 
    transforming the  ansatz wavefunction by 
    $\lvert \psi_{{\rm ansatz}} \rangle \to e^{i \theta_i \hat{O}_i}\lvert \psi_{{\rm ansatz}} \rangle$,
    for an arbitrary parameter $\theta_i$.
    \item[4.] 
    Identify the operator, $\hat{O}_n$, with the largest magnitude commutator with the Hamiltonian.
    If the absolute value of this commutator is below some pre-determined threshold, terminate the algorithm. 
    If it is above the threshold, 
    update the ansatz with the parameterized evolution of the operator $\lvert \psi_{{\rm ansatz}} \rangle \to e^{i \theta_n \hat{O}_n}\lvert \psi_{{\rm ansatz}} \rangle$.
    \item[5.] Use VQE to find the values of the variational parameters that minimize the energy,  \\
    $\langle\psi_{{\rm ansatz}} (\theta_{1}, \theta_{2},..., \theta_n)\rvert \hat{H}\lvert\psi_{{\rm ansatz}} (\theta_{1}, \theta_{2},..., \theta_n)\rangle$.
    The previously optimized values for $\theta_{1,2,...,n-1}$
    and $\theta_n=0$, are used as initial conditions. 
    If the optimal value of the newest parameter, $\theta_n$, is below some pre-determined threshold, terminate the algorithm.
    \item[6.] Return to step 3.
\end{itemize}
For a given pool of operators, it is {\it a priori} 
unknown if this algorithm will furnish a wavefunction that satisfies the 
pre-determined threshold
for the observable(s) of interest,
but it is expected that the  pool can be expanded on the fly to 
achieve the desired threshold.
The systems that have been explored with this algorithm show,
for a fixed pool,
exponential convergence with increasing numbers of iterations~\cite{Grimsley_2019,Tang:2021PRX,Bertels:2022oga,Romero:2022blx,Feniou:2023gvo}.

Generally,
different terms contributing to operators
in the pool 
do not commute with each other.
Constructing quantum circuits that exactly implement the exponential of a sum of non-commuting terms is challenging, and in practice approximations such as first-order Trotterization are used.
This introduces (higher-order) systematic deviations from the target unitary operator in each case, and 
defines the pool of unitary operators,
\begin{align}
     \{ \hat{U}_i \} & = \{ {\exp}(i \theta_i \hat{O}_i) \}\rightarrow \left \{ \prod\limits_t \hat{U}_i^{(t)} \right \}
    \ .
    \label{eq:TrotOp}
\end{align}
These Trotterized unitary operators correspond to the quantum circuits 
that are implemented in state preparation.
In optimization of the quality of the state prepared on a given quantum computer, 
particularly a NISQ-era device,
there are tradeoffs between the gate-depth of a particular circuit implementation, 
the coherence time, 
the errors associated with gate operations, 
and the associated Trotter errors.
This is explored in App.~\ref{app:TrotterErrors}.

Typically, ADAPT-VQE is a hybrid classical-quantum algorithm that evaluates matrix elements of the Hamiltonian 
in trial wavefunctions on a quantum computer, with parameters that are optimized classically.
One disadvantage of this is that the evaluation of expectation values of the Hamiltonian requires a large number of measurements (shots) on quantum computers.
A novel part of SC-ADAPT-VQE is the use of a {\it classical} simulator to determine the ADAPT-VQE state preparation circuits.
As shown in Sec.~\ref{sec:vacum_class_quan}, these circuits can be scaled and used to prepare the vacuum on arbitrarily large lattices.

\subsection{A Scalable Operator Pool for the Lattice Schwinger Model}
\label{sec:CQAdaptVQESM}
\noindent
A successful application of SC-ADAPT-VQE to the preparation of the lattice Schwinger model vacuum requires choosing an efficient and scalable pool of operators (first step in Fig.~\ref{fig:SCADAPTVQE-diag}).
These operators are
used to systematically improve the ansatz vacuum wavefunction, 
and are (only) constrained to be
charge neutral, 
symmetric under charge-conjugation and parity (CP) and, as a consequence of the CPT theorem~\cite{Bell:1955djs,Schwinger:1951xk,Luders:1954zz}, invariant under time reversal.\footnote{In the total charge $Q=0$ sector, there is a CP symmetry corresponding to the composition of a reflection through the mid-point of the lattice, exchanging {\it spatial} sites $n \leftrightarrow L-1-n$, and an interchange of an electron and a positron on each spatial site. 
In terms of spins on {\it staggered} sites this is realized as $\hat{\sigma}^i_n \leftrightarrow \hat{\sigma}_{2L-1-n}^i$ followed by $\hat{\sigma}_n^i \leftrightarrow \hat{X}_n\hat{\sigma}_n^i\hat{X}_n$, where $\hat{\sigma}^i$ with $i=1,2,3$ are the Pauli matrices.
For example, under a CP transformation, the following $L=4$ state becomes
\begin{align}
&
|\uparrow\downarrow\ \uparrow\uparrow\ \downarrow\downarrow\ \downarrow\uparrow\rangle \ = \ |. .\ \ . e^-\ \ e^+ .\ \ e^+ e^-\rangle \ \xrightarrow[ ]{\text{CP}} 
\ |\downarrow\uparrow\ \uparrow\uparrow\ \downarrow\downarrow\ \uparrow\downarrow\rangle 
\ =\ 
\ |e^+ e^-\ \ . e^-\ \ e^+ .\ \ . .\rangle
\ .
\label{eq:CP}
\end{align}
}
Ideally one wants to find the smallest pool of operators 
that is expressive enough to converge rapidly toward the vacuum.
For a lattice with OBCs, 
the system has translational symmetry in the volume that is broken by the boundaries (surface).
In the vacuum, 
the effects of the boundaries are expected to be localized, 
with penetration depths set by the mass gap.
Therefore, 
the pool of operators should contain 
translationally invariant ``volume" operators, 
and ``surface" operators that have support only near the boundaries.
In addition, a hierarchy is anticipated in which one-body operators 
are more important 
than two-body operators, 
two-body more important than three-body, and so on.\footnote{An $n$-body operator involves $n$ fermionic creation and $n$ fermionic annihilation operators.}
Note that because wavefunctions are evolved with ${\exp}(i \theta_i \hat{O}_i)$, 
arbitrarily high-body correlations are built from $n$-body operators 
(analogous to connected vs disconnected Feynman diagrams).
For the Schwinger model, we observe that one-body operators are sufficient.

With the above discussion as guidance, 
it is convenient to define two classes of one-body operators, 
one containing volume operators, 
and the other containing surface operators: 
\begin{align}
\hat{\Theta}_m^V &=   \frac{1}{2}\sum_{n=0}^{2L-1} (-1)^{n} \hat Z_n \ , \nonumber \\
\hat{\Theta}_{h}^V(d) &= \frac{1}{4}\sum_{n=0}^{2L-1-d} \left (\hat X_n \hat Z^{d-1} \hat X_{n+d} 
+ \hat Y_n \hat Z^{d-1} \hat Y_{n+d} \right ) \ ,  \nonumber \\
\hat{\Theta}_{m}^S(d) &= (-1)^d \frac{1}{2}\left ( \hat Z_d - \hat Z_{2L-1-d} \right ) 
\ , \nonumber \\
\hat{\Theta}_{h}^S(d) &= \frac{1}{4}\left (\hat X_1\hat Z^{d-1}\hat X_{d+1} + \hat Y_1\hat Z^{d-1}\hat Y_{d+1} 
\ + \  \hat X_{2L-2-d}\hat Z^{d-1}\hat X_{2L-2} + \hat Y_{2L-2-d}\hat Z^{d-1}\hat Y_{2L-2}\right )
\ .
\label{eq:PoolJW}
\end{align}
Unlabelled $\hat Z$s 
act on the qubits between the
leftmost and rightmost operators
(e.g., $\hat X_0 \hat Z^2 \hat X_3 = \hat X_0 \hat Z_1 \hat Z_2 \hat X_3$).
The first two operators in Eq.~(\ref{eq:PoolJW}) are translationally invariant,
$\hat{\Theta}_m^V$ is the mass term in the Hamiltonian,
and 
$\hat{\Theta}_{h}^V(d)$ is a generalized hopping term that spans an odd-number of fermion sites, $d$, connecting electrons and positrons at spatial sites separated by $\Delta L = (d-1)/2$. 
Only $d$-odd operators are retained, as the $d$-even operators break CP.
The second two operators in Eq.~(\ref{eq:PoolJW})
correspond to surface terms, of the form of a mass-density 
and of a hopping-density at and near the boundaries.
For $\hat{\Theta}_h^V(d)$, $d\in \{1,3,\ldots 2L-3\}$, and for  $\hat{\Theta}_{h}^S(d)$, $d\in \{1,3,\ldots 2L-5\}$,
preventing hopping between boundaries (which is found to improve convergence).

Time reversal symmetry implies that the vacuum wavefunction 
can be made real up to an overall phase. 
The SC-ADAPT-VQE ansatz is built from unitaries of the form $e^{i \theta_i \hat{O}_i}$, and furnishing a real wavefunction requires that the 
operators in the pool
are imaginary and anti-symmetric.
The operators in Eq.~(\ref{eq:PoolJW}) are real and are therefore disqualified from being members of the pool.
Instead, consider a pool comprised
of their commutators,\footnote{The commutators of $\hat{\Theta}$ operators not included in the pool are linear combinations of those that are.}
\begin{align}
\{ \hat{O} \} &= \left \{ \hat O_{mh}^{V}(d) \ , \ \hat O_{mh}^{S}(0,d) \ , \ \hat O_{mh}^{S}(1,d)
\right \}
\ ,\nonumber\\
 \hat O_{mh}^{V}(d) & \equiv i\left [\hat{\Theta}_m^V, \hat{\Theta}_{h}^V(d)\right ]
 = 
\frac{1}{2}\sum_{n=0}^{2L-1-d}(-1)^n\left (
\hat X_n\hat Z^{d-1}\hat Y_{n+d} - 
\hat Y_n\hat Z^{d-1}\hat X_{n+d} 
\right )
\ ,\nonumber\\
\hat O_{mh}^{S}(0,d) & \equiv i\left [\hat{\Theta}_{m}^S(0), \hat{\Theta}_{h}^V(d) \right ]
 = 
\frac{1}{4}\left (\hat X_0\hat Z^{d-1}\hat Y_{d} - \hat Y_0\hat Z^{d-1}\hat X_{d} 
- \hat Y_{2L-1-d}\hat Z^{d-1}\hat X_{2L-1} + \hat X_{2L-1-d}\hat Z^{d-1}\hat Y_{2L-1}\right ) 
\ ,\nonumber \\
 \hat O_{mh}^{S}(1,d) & \equiv i\left [\hat{\Theta}_{m}^S(1), \hat{\Theta}_{h}^S(d) \right ]
 = 
\frac{1}{4}\left (\hat Y_1\hat Z^{d-1}\hat X_{d+1} - \hat X_1\hat Z^{d-1}\hat Y_{d+1} 
+ \hat Y_{2L-2-d}\hat Z^{d-1}\hat X_{2L-2} - \hat X_{2L-2-d}\hat Z^{d-1}\hat Y_{2L-2} \right )
\ . 
\label{eq:poolComm}
\end{align}
While the contributions to extensive quantities from the 
volume operators,
$\hat O^{V}$,
typically scale as ${\cal O}(L)$,  the surface operators, $\hat O^{S}$,
make ${\cal O}(1)$ contributions as they are constrained to regions near the boundaries.\footnote{For the range of $m$ and $g$ we have considered, it was only necessary to consider $\hat{\Theta}_{m}^S(d)$ with $d=0,1$ in the pool.
Taking the continuum limit, where the correlation length diverges, will likely require keeping terms with $d>1$.} 
When acting on the strong-coupling vacuum, 
the exponential of an operator in the pool creates and annihilates $e^+ e^-$ pairs separated 
by distance $d$.
As the operators that are being considered are one-body, 
the variational algorithm is essentially building a coupled cluster singles (CCS) state (see, e.g., Refs.~\cite{RevModPhys.79.291,Hagen:2013nca}).

\section{Scalable Quantum Circuits from Classical Computing}
\label{sec:ClassSim}
\noindent
Integral to the application of SC-ADAPT-VQE 
is performing ADAPT-VQE on a series of systems that are large enough to enable a robust scaling of the parameterized circuits.
These scalable circuits can either be determined with classical computing, or by use of a smaller partition of a larger quantum computer.
In this section, 
SC-ADAPT-VQE is implemented using the ${\tt qiskit}$ noiseless classical simulator~\cite{IBMQ,qiskit}.

\subsection{Trotterized Quantum Circuits for the Scalable Operator Pool}
\noindent
As  discussed above, implementing the unitary operators in the pool, 
i.e., Eq.~\eqref{eq:TrotOp},
on classical simulators or quantum computers
requires mapping them to sequences of quantum gates.
For the individual terms in 
Eq.~\eqref{eq:poolComm}, 
we have chosen to do this using Trotterization.
The optimal gate decomposition 
is less important for implementation using a classical simulator,
but is crucial for successful simulations on a quantum computer.
With the goal of 
using IBM's superconducting-qubit quantum computers~\cite{IBMQ,qiskit},
our circuit designs aim to minimize two qubit gate count and circuit depth and require only nearest-neighbor connectivity.

\begin{figure}[t!]
    \centering
    \includegraphics[width=0.75\columnwidth]{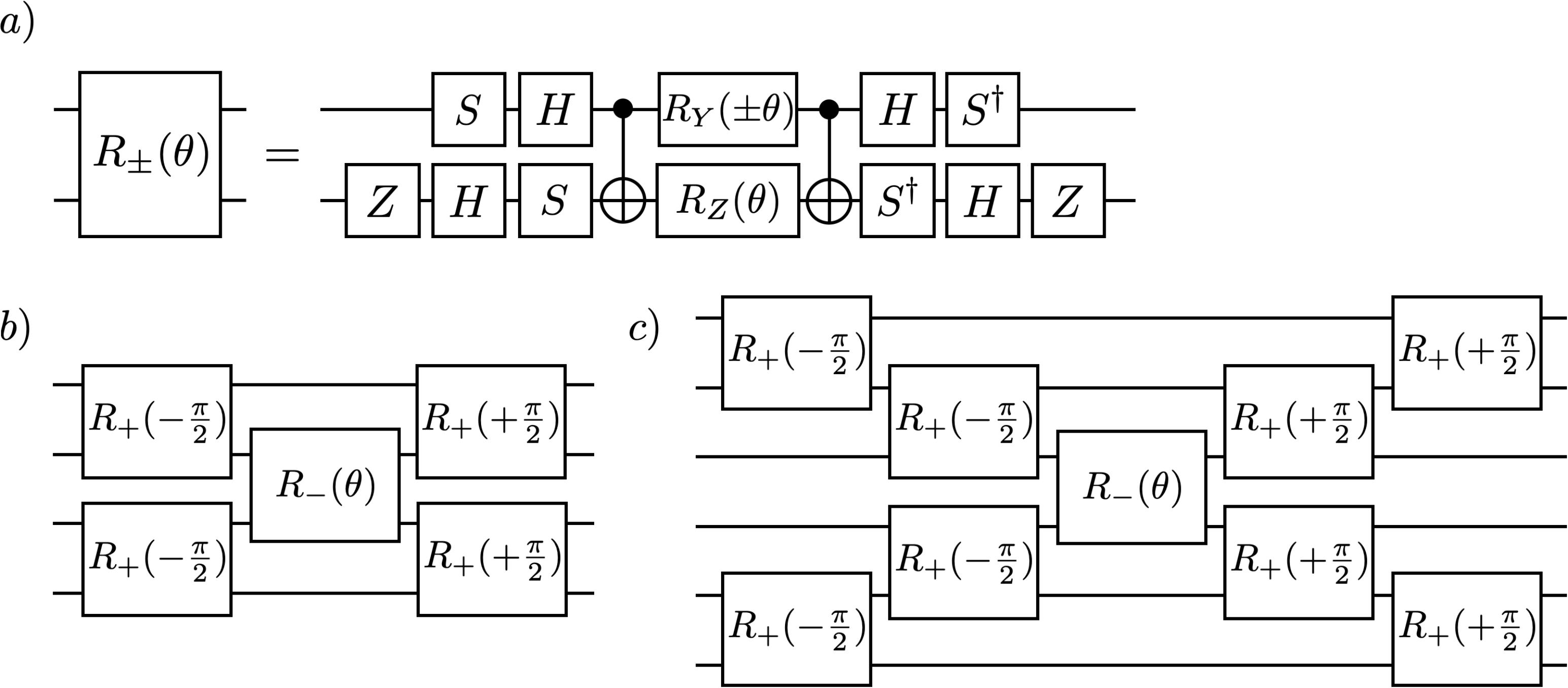}
    \caption{(a) Definition of the $R_{\pm}(\theta)$ gate, which
    implements ${\rm exp}[-i\theta/2 (\hat{Y}\hat{X}\pm \hat{X}\hat{Y})]$. 
    The $R_{\pm}(\theta)$ gate is used to implement 
    (b) ${\exp}[-i \theta/2 (\hat X\hat Z^2\hat Y - \hat Y\hat Z^2\hat X)]$ 
    and (c) ${\exp}[i \theta/2 (\hat X\hat Z^4\hat Y - \hat Y\hat Z^4\hat X)]$ (note the change in sign).}
    \label{fig:RpmOh35circ}
\end{figure}

\begin{figure}[t!]
    \centering
    \includegraphics[width=0.8\columnwidth]{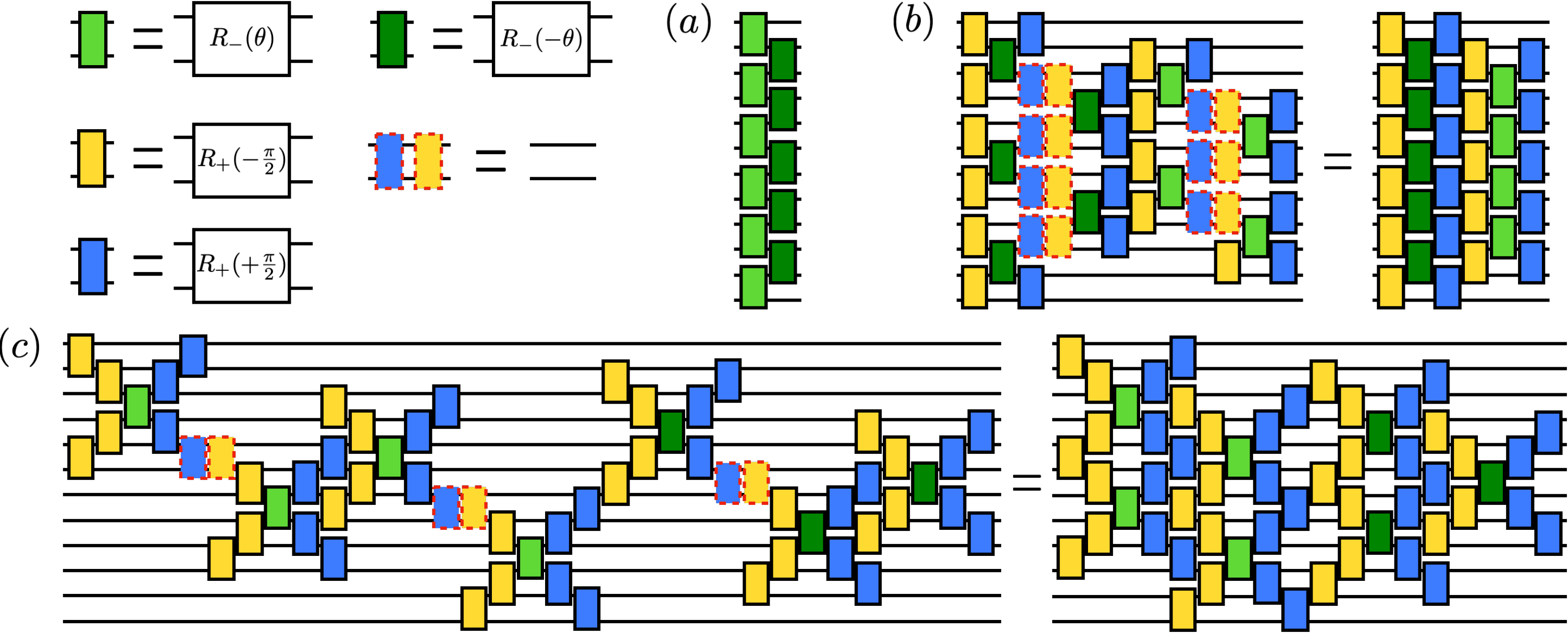}
    \caption{Simplifications of quantum circuits for the Trotterized unitaries corresponding to
    (a) $\hat O_{mh}^{V}(1)$, 
    (b) $\hat O_{mh}^{V}(3)$, 
    and (c) $\hat O_{mh}^{V}(5)$ for $L=6$, as explained in the main text. 
    Cancellations between $R_{+}(\pm \tfrac{\pi}{2})$ are highlighted with red-dash-outlined boxes.}
    \label{fig:ohd35multicirc}
\end{figure}

As can be seen in Eq.~\eqref{eq:poolComm}, 
each term in a given operator in the pool is of the form 
$(\hat{X} \hat{Z}^{d-1} \hat{Y} - \hat{Y} \hat{Z}^{d-1} \hat{X})$ 
for some odd value of $d$.
The construction of circuits implementing the corresponding unitary operators  follows the strategy outlined in Ref.~\cite{Algaba:2023enr}.
First, consider the Trotterization of terms with $d=1$, 
i.e., constructing a circuit corresponding to 
$e^{i \theta/2 (\hat{X}\hat{Y} \pm \hat{Y}\hat{X})} \equiv R_{\pm}(\theta) $.
There is a known 2-CNOT realization of this unitary operator~\cite{Algaba:2023enr}, 
shown in Fig.~\ref{fig:RpmOh35circ}a.
For terms with $d>1$, this circuit can be extended in an ``X" pattern as shown in Fig.~\ref{fig:RpmOh35circ}b and~\ref{fig:RpmOh35circ}c for $d=3$ and $d=5$, respectively.\footnote{
These circuits have been verified by comparison with 
Trotterized exponentials of fermionic operators.}
Terms with larger $d$ are constructed by extension of the legs of the ``X".
Compared with the traditional CNOT staircase-based circuits, 
there is a reduction by two CNOTs, 
and a reduction by a factor $\times 2$ in CNOT-depth.\footnote{The staircase circuit can be modified into an X-shaped one, reducing the depth, but with the same number of CNOTs~\cite{Cowtan:2019}.}
However, the primary advantage of these circuits is that they allow for an efficient arrangement of terms leading to cancellations among neighboring $R_+(\pm\tfrac{\pi}{2})$ gates.
As depicted in Fig.~\ref{fig:ohd35multicirc}, this is made possible by arranging the circuit elements so that sequential terms are offset by $d-1$ qubits, 
i.e., start on qubit $\{0,d-1,2(d-1),\ldots\}$.
This allows the outermost gates to cancel (using the identity in the upper left of Fig.~\ref{fig:ohd35multicirc}). 
Also, for $d\geq 5$, the next layer should start $(d-1)/2$ qubits below the previous one, 
as the circuit depth can be reduced
by interleaving the legs of the ``X".
Further optimizations are possible by noting that distinct orderings of terms, while equivalent up to higher order Trotter errors, can have different convergence properties; see App.~\ref{app:TrotterErrors}.
%

\subsection{
Building Scalable State Preparation Quantum Circuits using SC-ADAPT-VQE with Classical Computing
}
\noindent
In this section, SC-ADAPT-VQE is used to prepare approximations to the vacuum of the lattice Schwinger model on up to $L=14$ spatial sites (28 qubits) using classical simulations of the quantum circuits developed in the previous section (second step in Fig.~\ref{fig:SCADAPTVQE-diag}).

%
\begin{table}[!ht]
\renewcommand{\arraystretch}{1.4}
\begin{tabularx}{\textwidth}{|c || Y | Y || Y | Y || Y ||  Y |}
 \hline
 $L$ 
 & $\varepsilon^{\rm (aVQE)}$ 
 & $\varepsilon^{\rm (exact)}$ 
 & $\chi^{\rm (aVQE)}$ 
 & $\chi^{\rm (exact)}$ 
 & $\infiL$ & \text{\# CNOTs/qubit} 
  \\
 \hline\hline
 6 & -0.30772& -0.30791  & 0.32626& 0.32720 & 0.00010 & 31.2\\
 \hline
7 & -0.31097& -0.31117  & 0.32847& 0.32947 & 0.00011 & 33.6\\
\hline
8 & -0.31348& -0.31363  & 0.33036& 0.33118 & 0.00008 & 35.8\\
\hline
9 & -0.31539& -0.31553  & 0.33171& 0.33251 & 0.00008 & 37.1\\
\hline
10 & -0.31691& -0.31706  & 0.33279& 0.33358 & 0.00008 & 38.2\\
\hline
11 & -0.31816& -0.31831  & 0.33367& 0.33445 & 0.00008 & 39.1\\
\hline
12 & -0.31920& -0.31935  & 0.33441& 0.33517 & 0.00008 & 39.8\\
\hline
13 & -0.32008& -0.32023 &  0.33504& 0.33578 & 0.00008 & 40.5\\
\hline
14 & -0.32084 & -0.32098 & 0.33557 & 0.33631 &0.00008 & 41.0\\
 \hline
\end{tabularx}
\caption{
Energy density, chiral condensate and wavefunction infidelity 
for the vacuum of the Schwinger model with $m=0.5,g=0.3$.
Both the results obtained from 7 steps of the SC-ADAPT-VQE (aVQE) algorithm using {\tt qiskit}'s classical simulator and the exact values
are given.
The last column shows the number of CNOTs/qubit in the state preparation circuit.}
 \label{tab:Echifmp5gp3}
\end{table}
\begin{figure}[!ht]
    \centering
    \includegraphics[width=\columnwidth]{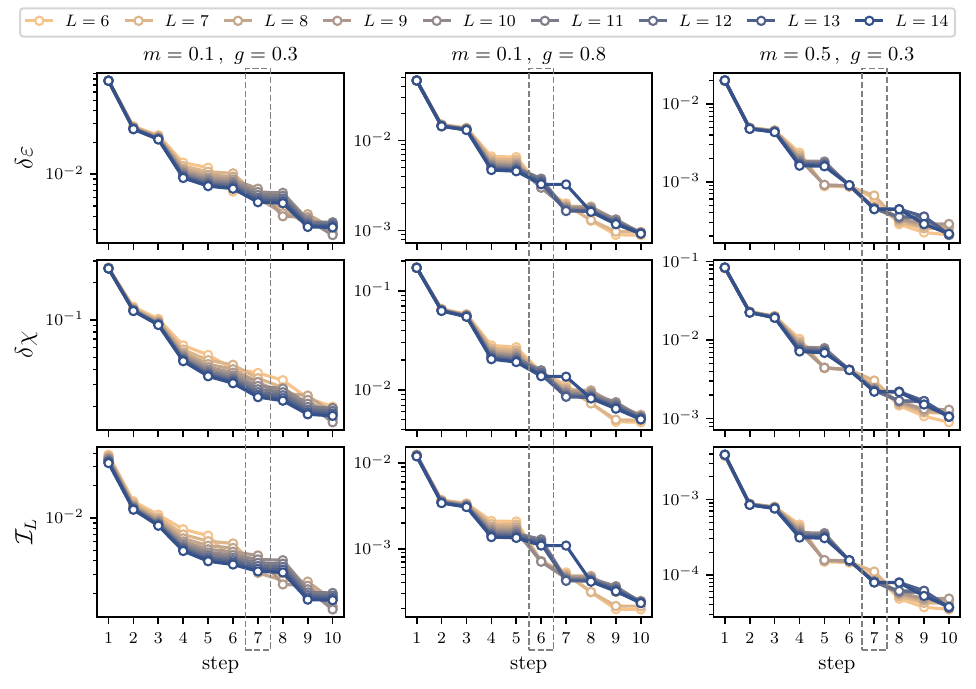}
    \caption{Deviations from the exact values of the energy density $\delta \varepsilon$, chiral condensate $\delta \chi$, and 
    wavefunction infidelity density $\infiL$ 
    obtained with SC-ADAPT-VQE. 
    The deviation in quantity ``$x$'' is defined 
    as $\delta x = |(x^{(\rm aVQE)}-x^{(\rm exact)})/x^{(\rm exact)}|$,
    where $x^{(\rm exact)}$ denotes the exactly calculated value at the same $L$.    
    Results are shown for $L=6$ to $L=14$   
    as a function of step number for $m=0.1, g=0.3$ (left column),  
    $m=0.1, g=0.8$ (center column) 
    and $m=0.5, g=0.3$ (right column).
    The numerical values for $m=0.5, g=0.3$ for the $7^{\rm th}$ step (highlighted with the dashed box) are given in Table~\ref{tab:Echifmp5gp3}, and the sequencing  of the corresponding Trotterized operators 
    and the variational parameters are given in Table~\ref{tab:AnglesXCircmp5gp3}.
    The corresponding results for $m=0.1,g=0.3$ ($7^{\rm th}$ step) and $m=0.1, g=0.8$ ($6^{\rm th}$ step) can be found in  App.~\ref{app:classData}.
    }
    \label{fig:ClassConvXCirc}
\end{figure}
\FloatBarrier
In addition to the energy density and chiral condensate introduced in Sec.~\ref{sec:infVolExtra}, the infidelity density,
\begin{equation}
\infiL = \frac{1}{L} \left( 1 - 
|\langle\psi_{\rm ansatz}|\psi_{\rm exact}\rangle|^2 \right)
\ ,
\label{eq:fiddens}
\end{equation} 
is also studied, where $|\psi_{\rm exact}\rangle$ is the exact vacuum wavefunction on a lattice with $L$ spatial sites.
An infidelity density that is constant in $L$ corresponds to constant deviations in local observables evaluated in the prepared state.

To investigate the interplay between $L$ and 
$\xi=1/m_{\text{hadron}}$, 
three sets of parameters are considered: 
$m=0.1, g=0.3$ ($\xi_{L=14} = 2.6$), 
$m=0.1, g=0.8$  ($\xi_{L=14} = 1.3$) 
and $m=0.5, g=0.3$  ($\xi_{L=14} = 0.9$).
The $\xi$ are determined with exact diagonalization, and are found to weakly depend on $L$.
Note that increasing either $m$ or $g$ decreases the correlation length.
To make systematically improvable predictions 
of observables from the QFT that emerges from a given lattice model, 
extrapolations to the continuum (lattice spacing to zero) and infinite-volume ($L \to \infty)$ limits must be performed.
This requires that the relevant correlation length(s) are all much greater than the lattice spacing, $\xi \gg 1$ in lattice units, but are well contained in the lattice volume, $L\gg\xi$.
We primarily focus on extrapolation to large lattices, and therefore only require $L\gg \xi$.
As a result, the parameter set $m=0.5, g=0.3$ is used as the primary example throughout this work.

The values of $\varepsilon$, $\chi$ and $\infiL$ obtained at 
the $7^{\rm th}$ step of SC-ADAPT-VQE with $m=0.5, g=0.3$ are given in 
Table~\ref{tab:Echifmp5gp3},
while their deviations from the exact values 
are shown in Fig.~\ref{fig:ClassConvXCirc},
as a function of increasing number of SC-ADAPT-VQE steps.
The corresponding numerical values obtained from the other parameter sets are presented in App.~\ref{app:classData}.\footnote{The $6^{\rm th}$ and $7^{\rm th}$ steps were chosen for study in detail as the operator ordering has stabilized for $L\leq 14$. This allows the operator structure to be displayed in a single table, and enables the systematic extrapolation of parameters. 
The available classical computing resources limited the maximum number of steps of SC-ADAPT-VQE to 10.}
As seen by their approximately linear behavior in the log-plots in Fig.~\ref{fig:ClassConvXCirc}, 
the error in each of these quantities decreases exponentially with algorithm step,
indicating convergence to the target wavefunction.
This exponential trend is demonstrated out to 10 steps, reaching a convergence comparable to the systematic errors introduced in the $L$-extrapolations below.
This provides evidence that this choice of initial state and operator pool does not suffer from ``barren plateaus" or local minima.
For a given step in the algorithm, 
the error is seen to become independent of system size.
This indicates that 
extrapolations of the circuits to arbitrarily large systems will have errors that are {\it independent of $L$}.
As discussed above, it is expected that SC-ADAPT-VQE will converge more rapidly for systems with smaller correlation lengths.
This is indeed seen in Fig.~\ref{fig:ClassConvXCirc}, 
where the correlation length decreases from left to right, while the convergence improves.
Also included in Table~\ref{tab:Echifmp5gp3} is the number of CNOTs per qubit in the SC-ADAPT-VQE circuit. 
It is seen to  scale as a constant plus a subleading ${\mathcal O}(1/L)$ term,
leading to an asymptotic value of 48 CNOTs per qubit.
This scaling is due to there being $(2L-d)$ terms in each volume operator.

The structure of the SC-ADAPT-VQE state preparation circuit and the corresponding variational parameters for $m=0.5$ and $g=0.3$ 
are given in Table~\ref{tab:AnglesXCircmp5gp3}.
Notice that initially localized operators are added to the wavefunction (small $d$), 
followed by increasingly longer-range ones, as well as surface operators. Systems with longer correlation lengths require larger $d$ operators (e.g., compare Table~\ref{tab:AnglesXCircmp5gp3} and Table~\ref{tab:AnglesXCircmp1gp3}), 
in line with previous discussions on the exponential decay of correlations for $d > \xi$.
It is also seen that the surface operators become less important 
(appear later in the ansatz structure) for larger lattices.
For example, as shown in Table~\ref{tab:AnglesXCircmp5gp3}, the $5^{{\rm th}}$ step of SC-ADAPT-VQE transitions from being a surface to a volume operator at $L=10$ (causing the jump in convergence at the fifth step in the right column of Fig.~\ref{fig:ClassConvXCirc}).
This is expected as they contribute ${\mathcal O}(1/L)$ 
to the energy density, whereas volume operators contribute ${\mathcal O}(1)$.

\begin{figure}[!b]
    \centering
    \includegraphics[width=0.65\columnwidth]{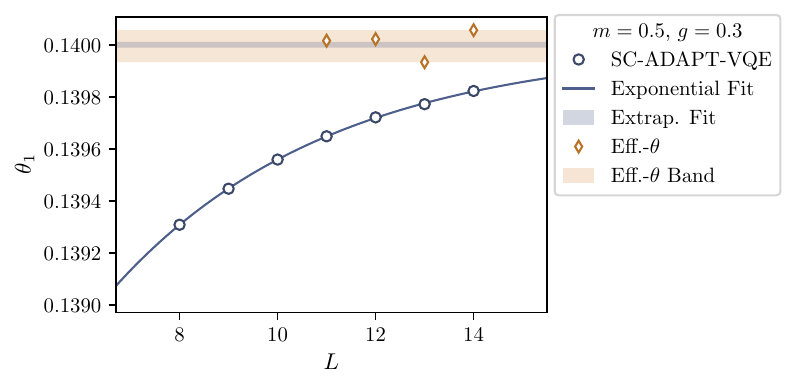}
    \caption{
    Example of fitting the asymptotic $L$-dependence of a parameter defining the SC-ADAPT-VQE state-preparation circuit.
    The results for $\theta_1$, corresponding to evolving by $\hat O_{mh}^{V}(1)$ (blue circles),   determined from classical simulations,
    for 
    $m=0.5$ and $g=0.3$ given in Table~\ref{tab:AnglesXCircmp5gp3},
    are extrapolated to $L=\infty$ by 
i) use of a 3-parameter fit given in Eq.~\eqref{eq:thetaextrap}, 
as shown by the blue line,
with an asymptotic value shown by the blue region,
and by
ii) the forming of effective-$\theta$ (orange diamonds) defined in Eq.~\eqref{eq:LdifsB},
with the maximum and minimum values shown as the orange shaded region.
}
    \label{fig:ThetaLextrap53}
\end{figure}
Importantly, Table~\ref{tab:AnglesXCircmp5gp3} shows that the order of operators, 
and the corresponding variational parameters are converging with increasing system size (third step in Fig.~\ref{fig:SCADAPTVQE-diag}). 
This is due to exponentially decaying correlations for $d \gg \xi$, 
and it is expected that the variational parameters will also converge exponentially,
once $L$ is sufficiently large to contain $\xi$, and we assume the following form:
\begin{equation}
\theta_i \ = \ \theta_i^{L=\infty} \ + \  c_1\,e^{-c_2 \, L} 
\ .
\label{eq:thetaextrap}
\end{equation}
Table~\ref{tab:AnglesXCircmp5gp3} shows that this convergence sets in 
for $L>7$,\footnote{The ordering of operators changes at $L=10$ but the operator content is unchanged, so it is still possible to use $L=8,9$ in the extrapolation.} 
and the variational parameters extrapolated to $L=\infty$ are given in the last row of Table~\ref{tab:AnglesXCircmp5gp3}.
These are used in the next section to initialize the vacuum on 
lattices up to $L=500$.
An example of 
extrapolating the variational parameters
is shown in Fig.~\ref{fig:ThetaLextrap53} for 
the parameter $\theta_1$,
associated with
$\hat O_{mh}^{V}(1)$.
The exact results obtained for $L\le 14$ are well reproduced and 
extrapolated with the exponential functional form in Eq.~(\ref{eq:thetaextrap}).\footnote{
One could imagine generating the $\theta^{L=\infty}_i$ for a variety of $m$ and $g$, and then machine learning the variational parameters for all $m$ and $g$.
This could be particularly useful for $m$ and $g$ that approach the continuum limit, where the correlation length can no longer be contained within lattice volumes accessible to classical simulators.}
A more complete discussion of the parameter extrapolations, along with
examples for $m=0.1$ and $g=0.3$ and for $m=0.1$ and $g=0.8$, can be found in App.~\ref{app:ThetaScaling}.

\begin{table}[!t]
\renewcommand{\arraystretch}{1.4}
\begin{tabularx}{\textwidth}{|c || Y | Y | Y | Y || Y |  Y| Y | Y | Y |}
 \hline
 \diagbox[height=23pt]{$L$}{$\theta_i$} & $\hat O_{mh}^{V}(1)$
 & $\hat O_{mh}^{V}(3)$ & $\hat O_{mh}^{V}(5)$ & $\hat O_{mh}^{V}(1)$ & $\hat O_{mh}^{V}(7)$ &  $\hat O_{mh}^{S}(0,1)$ & 
 $\hat O_{mh}^{V}(7)$  &
 $\hat O_{mh}^{V}(1)$ &
 $\hat O_{mh}^{S}(0,3)$ \\
 \hline\hline
 6 & 0.18426 & -0.03540 & 0.00731 & 0.11866 & -- & 0.06895 & -0.00182 & -- & -0.03145\\
 \hline
 7 & 0.18440 & -0.03574 & 0.00729 & 0.11864 & -- & 0.06867 & -0.00177 & -- & -0.03066\\
 \hline
 8 & 0.13931 & -0.03727 & 0.00760 & 0.08870 & -- & 0.06925 & -0.00183 & 0.07457 & --\\
 \hline
 9 & 0.13945 & -0.03714 & 0.00755 & 0.08849 & -- & 0.06904 & -0.00180 & 0.07473 & --\\
 \hline
 10 & 0.13956 & -0.03703 & 0.00752 & 0.08832 & -0.00178 & 0.06888 & -- & 0.07485 & --\\
 \hline
 11 & 0.13965 & -0.03695 & 0.00749 & 0.08819 & -0.00177 & 0.06875 & -- & 0.07494  & --\\
 \hline
 12 & 0.13972 & -0.03688 & 0.00747 & 0.08808 & -0.00176 & 0.06865 & -- & 0.07502 & -- \\
 \hline
 13 & 0.13977 & -0.03683 & 0.00745 & 0.08800 & -0.00175 & 0.06856 & -- & 0.07508 & --\\
 \hline
14&0.13982& -0.03678&0.00744&0.08793&-0.00174&0.06849&--&0.07513&--\\
\hline
 \hline
  $\infty$ &0.1400 & -0.0366 & 0.0074 & 0.0877 & -0.0017 & 0.0682 & -- & 0.0753 & --\\
 \hline
\end{tabularx}
\caption{Structure of the ansatz wavefunction with $m=0.5$ and $g=0.3$ 
through 7 steps of the SC-ADAPT-VQE algorithm 
obtained from a classical simulation using {\tt qiskit}. 
For a given $L$, the order that the operators are added to the ansatz is displayed from
left to right, with the associated parameter, $\theta_i$, given as the entry in the table.
The  operators,
$\hat O_{mh}^{V}(d_h)$ and $\hat O_{mh}^{S}(d_m,d_h)$,
are defined in Eq.~(\ref{eq:poolComm}).
An entry of  `` -- '' means that the operator does not contribute.
The bottom row corresponds to an extrapolation to $L=\infty$ as detailed in Eq.~\eqref{eq:thetaextrap}.
}
 \label{tab:AnglesXCircmp5gp3}
\end{table}
%

\section{Preparing the Vacuum of the Schwinger Model on Large Lattices}
\label{sec:vacum_class_quan}
\noindent
The vacuum preparation circuits, determined for $L\leq14$ with SC-ADAPT-VQE using an exact (statevector) classical simulator, are scaled to prepare the vacuum on much larger lattices. 
These scaled circuits are used to prepare the vacuum on lattices of up to $L=500$ (1000 qubits) using a classical MPS circuit simulator
and up to $L=50$ (100 qubits) using IBM's {\tt Eagle}-processor quantum computers (fourth step in Fig.~\ref{fig:SCADAPTVQE-diag}).
We emphasize that this scaling requires no further optimization of the circuits.
The chiral condensate and energy density determined from the classical simulator
are found to be consistent with DMRG calculations.
On the quantum computers, the chiral condensate and charge-charge correlators 
are measured
to probe the quality of one- and two-qubit observables.
The results are in agreement with those from the classical MPS simulator, within statistical uncertainties.

\subsection{
Classical Simulation}
\label{sec:scaling}
\noindent
Very large quantum circuits that do not generate long-range entanglement can be efficiently simulated using the 
{\tt qiskit} {\tt matrix\_product\_state} classical simulator.
Here it is used to simulate the preparation of the Schwinger model vacuum 
on $L\gg 14$ lattices,
applying the scalable circuits 
determined in the previous section
from 7 steps of SC-ADAPT-VQE on $L\le 14$ lattices.
The values obtained for the chiral condensate and energy density up to $L=500$ are compared with DMRG results, and are presented in Table~\ref{tab:SQCResults}.
The deviations in the energy density ($\sim 1\times 10^{-4}$) and chiral condensate ($\sim 1\times 10^{-3}$) are in good agreement with what was found for smaller $L$; 
see Table~\ref{tab:Echifmp5gp3}.
This demonstrates that the systematic errors in the vacuum wavefunctions prepared with the scaled quantum circuits are (approximately) independent of $L$ over this range of lattice volumes.\footnote{The $6^{{\rm th}}$ operator in the extrapolation is a surface operator, whose contribution to the energy density scales as $1/L$. 
Therefore, if SC-ADAPT-VQE could be performed on, for example, $L=500$, this operator would likely not be in the ansatz.
Evidently the ``error" introduced by extrapolating the ansatz with a surface operator is small since the deviation of observables for large $L$ is the same as for $L\leq14$.}
The scaled circuits corresponding to $m=0.1, g=0.3$ and $m=0.1, g=0.8$ have also been used to successfully prepare the vacuum.
However, due to the larger correlation lengths, 
MPS calculations with $L\gtrsim 100$ required excessive classical resources, and were not performed.  See App.~\ref{app:classData} for more details.
\begin{table}[!t]
\renewcommand{\arraystretch}{1.4}
\begin{tabularx}{\textwidth}{|c || Y | Y || Y | Y |}
 \hline
 $L$ 
 & $\varepsilon^{\rm (SC-MPS)}$ 
 & $\varepsilon^{\rm (DMRG)}$ 
 & $\chi^{\rm (SC-MPS)}$  
 & $\chi^{\rm (DMRG)}$ \\
 \hline\hline
 50 & -0.32790 & -0.32805 &0.34044 & 0.34123\\
 \hline
 100 & -0.32928 & -0.32942 & 0.34135 & 0.34219\\
 \hline
200  & -0.32996 & -0.33011 & 0.34181 & 0.34267 \\
\hline
300 & -0.33019 & -0.33034 & 0.34196 & 0.34282 \\
\hline
400 & -0.33031 & -0.33045 & 0.34204 &0.34291\\
\hline
500 & -0.33038 & -0.33052 & 0.34209 & 0.34296 \\
 \hline
\end{tabularx}
\caption{
Results for large lattices with $m=0.5, g=0.3$ through 7 steps of SC-ADAPT-VQE 
using circuits scaled from $L\leq14$.
The superscript ``SC-MPS'' denotes the results obtained 
from the scaled circuits
using the {\tt qiskit} MPS classical simulator,
and the superscript ``DMRG'' denotes results obtained from DMRG calculations.
}
 \label{tab:SQCResults}
\end{table}

It is worth summarizing what has been accomplished 
in this work with classical simulations:
\begin{itemize}
    \item In Sec.~\ref{sec:SchwingerHam}, the vacuum energy density and chiral condensate were determined exactly for 
    $L\le 14$ (28 staggered lattice sites) using exact diagonalization, and for $L\le 10^3$ using DMRG.
    The results for $L\ge 9$ were (consistently) extrapolated to $L\rightarrow\infty$, with $1/L$ scaling.
    \item In Sec.~\ref{sec:ClassSim}, SC-ADAPT-VQE, based on the scalable operator pool determined in Sec.~\ref{sec:II}, was performed on $L\leq 14$ lattices.
    Intensive quantities were found to converge exponentially with circuit depth, and the errors in these quantities, as well as the structure of the state preparation circuits, were found to become independent of $L$.
    This enabled the variational parameters defining the state preparation circuits to be consistently extrapolated to arbitrarily large $L$.
    \item 
    In this section,
    the quantum circuits corresponding to 7 steps of SC-ADAPT-VQE were scaled and applied to large lattices using the {\tt qiskit} MPS circuit simulator.  
    The deviations of the energy density and chiral condensate computed from these wavefunctions were found to be independent of $L$, i.e., consistent with $L\le 14$.
\end{itemize}
These main points indicate that the quantum circuits determined classically with SC-ADAPT-VQE can be used to prepare the vacuum 
of the Schwinger model 
on quantum computers at scale with a precision that is independent of system size.

\subsection{Quantum Simulations on 100 Qubits using IBM's Quantum Computers}
\label{sec:SCQuSim}
\noindent
The quantum circuits determined via classical simulation on $L\le 14$ lattices  
are now scaled to larger $L$ to
prepare the vacuum of the Schwinger model 
on up to 100 qubits of 
IBM's 127 superconducting-qubit 
{\tt Eagle} quantum computers with heavy-hexagonal communication fabric.
Hamiltonian parameters $m=0.5, g=0.3$ with $L=14,20,30,40,50$,
and 
state preparation circuits scaled from 2 steps of SC-ADAPT-VQE 
(compared to 7 steps in the previous section), are used.
Fewer steps equates to shallower circuits, and a preliminary study of the performance of the computer with more steps can be found in App.~\ref{app:qusimDetail}.
The variational parameters extrapolated to the chosen range of $L$ for 2 steps of SC-ADAPT-VQE are given in 
Table~\ref{tab:angles_ibm} in
App.~\ref{app:qusimDetail}.

The large number of qubits and two-qubit gates involved in these 
simulations make error mitigation essential to obtain reliable estimates of observables.
Specifically, this work uses readout-error mitigation (REM), dynamical decoupling (DD), Pauli twirling (PT), and decoherence renormalization. 
The {\tt qiskit} Runtime Sampler primitive is used to obtain readout-corrected quasi-distributions via the matrix-free measurement mitigation (M3) from Ref.~\cite{Nation:2021kye}. 
Also included in the primitive is DD, which is used to suppress crosstalk and idling errors~\cite{Viola:1998dsd,2012RSPTA.370.4748S,Ezzell:2022uat}.
Crucial to the error mitigation is decoherence renormalization~\cite{Urbanek_2021,ARahman:2022tkr,Farrell:2022wyt,Ciavarella:2023mfc}, 
modified in this work for simulations on a large number of qubits, which we call {\it Operator Decoherence Renormalization} (ODR).
Underpinning decoherence renormalization is PT~\cite{Wallman:2016nts}, which turns coherent two-qubit gate errors into incoherent errors, which can be inverted to recover error-free expectation values.
Unlike previous applications of decoherence renormalization, 
which assume a constant decoherence across the device, 
ODR estimates the decoherence separately for each operator.
This is done by running a mitigation circuit, which has the same 
operator structure as the one used to extract the observables, but with the noise-free result being known {\it a priori}. 
We choose the state preparation circuits with the variational parameters 
set to zero for mitigation,
and 
in the absence of noise this prepares the strong coupling vacuum, $\lvert \Omega_0 \rangle$ in Eq.~\eqref{eq:SCvac}.
Naively, it could be expected that post-selecting results on states with 
total charge $Q = 0$ would eliminate the leading bit-flip errors~\cite{Klco:2019evd}.
However, when post-selection is combined with ODR, which accommodates single-qubit decoherence, undesirable correlations between qubits are introduced.
We find that performing both mitigation techniques 
(post-selection and ODR)
degrades the quality of two-qubit observables, 
and post selection is not used in this work as it is found to be less effective.
More details about ODR and post-selection can be found in App.~\ref{app:QSimError}.

\begin{figure}[!t]
    \centering
    \includegraphics[width=\columnwidth]{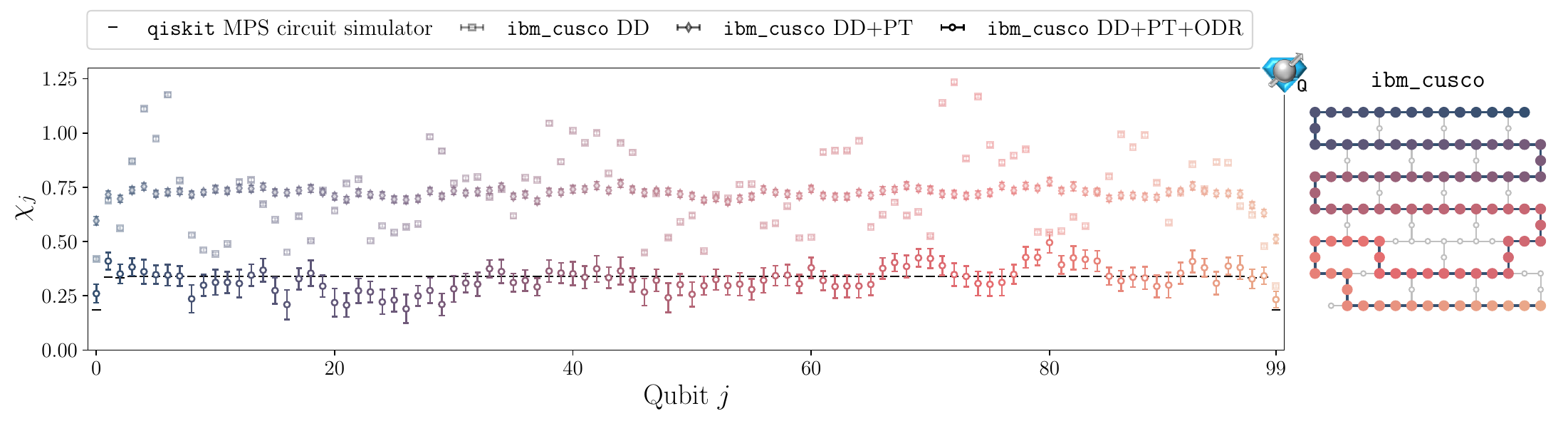}
    \caption{Local chiral condensate $\chi_j$ for $L=50$, as obtained from 
    IBM's Eagle-processor quantum computer
    {\tt ibm\_cusco} after different steps of error mitigation: DD (squares), PT (diamonds), and ODR (circles).
    This is compared with the expected results obtained from the {\tt qiskit} MPS circuit simulator (black dashes). 
    Averaging $\chi_j$ over all of the  
    qubits (including at the boundaries) gives the chiral condensate presented in Table~\ref{tab:SQCResults_ibm}.
    The layout of the qubits used on the processor is shown on the right.
    These results were obtained by performing 150 Pauli twirls, each involving $8\times 10^3$ shots
    for the physics circuits and the corresponding mitigation circuits.
    The blue icon in the upper right indicates that this calculation was done on a quantum device~\cite{Klco:2019xro}.}
    \label{fig:chi_50}
\end{figure}

The local chiral condensate, $ \chi_j $ in Eq.~(\ref{eq:chiralCond}), 
obtained from {\tt ibm\_cusco} for $L=50$ is shown in Fig.~\ref{fig:chi_50},
where the subscript ``$j$" denotes the qubit index.\footnote{
For all of the results presented in this work, 
correlated bootstrap re-sampling was used to estimate statistical (shot) uncertainties.
The circuits used for $L\leq 40$ were executed on {\tt ibm\_brisbane} with 
40 Pauli-twirled instances for both the mitigation and the physics circuits, 
each with $8\times10^3$ shots. 
For $L=50$, the M3 method was not applied due to the large overhead in classical computing, 
and production was executed on {\tt ibm\_cusco} with 150 Pauli-twirled instances.
Additional details can be found in App.~\ref{app:qusimDetail}.}
Three different sets of results (in different stages of error mitigation) are shown: with only DD applied (squares), with DD and PT applied (diamonds), and after ODR (circles). Looking at the results with only DD (squares), it is seen that the noise is not uniform across the device, signaling a significant contribution of coherent noise. After PT (diamonds), this coherent noise is averaged out, and is transformed into incoherent (depolarizing) noise, seen by the almost-constant shift of the results compared with the MPS simulation. Finally, ODR removes this shift by mitigating the effects of depolarizing noise. More details on the interplay between these methods can be found in App.~\ref{app:qusimDetail}.

With the statistics and twirlings gathered,
the $1\sigma$ uncertainties in each point are $\sim 15\%$ of their mean, 
and each $ \chi_j $ is within $3 \sigma$ of the MPS simulator result (the individual values of $\chi_j$ can be CP averaged to reduce the uncertainty, as shown in Fig.~\ref{fig:chi_14-50_CP} in App.~\ref{app:qusimDetail}).
It is expected that these uncertainties will reduce with increased statistics 
and twirlings.
Notice that the expected values of 
$ \chi_j $ deviate from the volume average 
for only a few qubits near the boundaries.
This is because the boundaries  perturb the wavefunction only over a few correlation lengths, leaving the rest of the volume unaffected.
The chiral condensates for $L=14,20,30,40$ and 50 are given in Table~\ref{tab:SQCResults_ibm}.
This is an average over the whole lattice, Eq.~\eqref{eq:chiralCond}, and therefore
the uncertainty decreases with increasing $L$
due to increased sampling.
Despite having smaller uncertainties, 
the results remain within $3\sigma$ of the MPS simulator result.
Also given in Table~\ref{tab:SQCResults_ibm} is the number of two-qubit CNOT gates.
The number of CNOTs is seen to grow linearly with $L$, without affecting the quality of the result, and 788 CNOTs over 100 qubits is well within the capabilities of the quantum computer.
This is in line with other quantum simulations that have been performed with large numbers of qubits and CNOTs using IBM's quantum computers~\cite{Yu:2022ivm,Kim:2023bwr,Shtanko:2023tjn}.
\begin{table}[!t]
\renewcommand{\arraystretch}{1.4}
\begin{tabularx}{\textwidth}{|c |c |c || Y | Y | Y | Y |}
 \hline
 $L$ & Qubits & CNOTs
 & $\chi^{\rm (SC-{\tt IBM})}$ before ODR 
 & $\chi^{\rm (SC-{\tt IBM})}$  after ODR
 & $\chi^{\rm (SC-MPS)}$
 & $\chi^{\rm (DMRG)}$ \\
 \hline\hline
 14 & 28 & 212 & 0.491(4)  & 0.332(8) & 0.32879 & 0.33631 \\
 \hline
 20 & 40 & 308 & 0.504(3) & 0.324(5) & 0.33105 & 0.33836 \\
 \hline
 30 & 60 & 468 & 0.513(2) & 0.328(4) & 0.33319 & 0.33996 \\
\hline
 40 & 80 & 628 & 0.532(2) & 0.334(3) & 0.33444 & 0.34075 \\
\hline
 50 & 100 & 788 & 0.721(2) & 0.326(3) & 0.33524 & 0.34123 \\
 \hline
\end{tabularx}
\caption{
Chiral condensate in the Schwinger model vacuum 
obtained from {\tt ibm\_brisbane} ($L\leq 40$) and {\tt ibm\_cusco} ($L=50$) for large lattices with $m=0.5, g=0.3$ using the scaled circuits from two steps of SC-ADAPT-VQE.
The values before and after application of ODR are given in columns four and five. 
Column six gives results obtained from running the two step SC-ADAPT-VQE circuits on an MPS classical simulator (the noiseless result), while column seven gives the results from DMRG calculations.}
\label{tab:SQCResults_ibm}
\end{table}
This highlights the fact that it is not the total number of CNOT gates in the
quantum circuit that is limiting the scale of simulations, but rather it is the number of CNOT gates per qubit. 
This, of course, assumes that the CNOT gates in a single layer of the circuit can be enacted in parallel.
Due to this, increasing $L$ actually improves volume-averaged quantities by $\sim 1/\sqrt{L}$ due to statistical averaging.
In a similar vein, since scalable circuits repeat structures of size $\xi$ many times over the whole lattice, 
the number of Pauli-twirls being sampled is effectively multiplied by $L/\xi$.

\begin{figure}[!t]
    \centering
    \includegraphics[width=0.95\columnwidth]{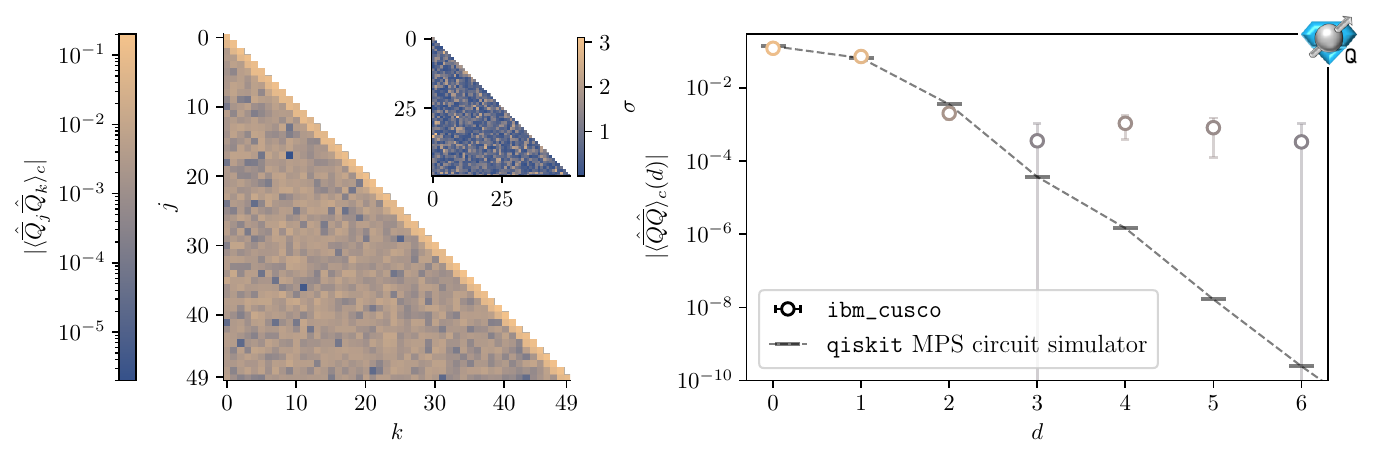}
    \caption{
    Left: Connected contributions to the spatial charge-charge correlation functions, 
    $\langle \hat{\overline{Q}}_{j} \hat{\overline{Q}}_{k} \rangle_c$, 
    for $L=50$ 
    (the inset shows the number of standard deviations by which the results obtained from {\tt ibm\_cusco} deviate from the MPS simulator results).
    Right: Volume averaged correlation functions as a function of distance $d$, 
    $\langle \hat{\overline{Q}} \hat{\overline{Q}} \rangle_c (d)$, with the points following the same color map as in the left main panel (error bars show $1\sigma$ standard deviations).}
    \label{fig:qiqj_40}
\end{figure}

To further probe the quality of the prepared wavefunctions, correlations between electric charges on the spatial sites are considered.
The charge on a spatial site is defined 
to be the sum of charges on the two associated staggered sites, 
$\hat{\overline{Q}}_{k} = \hat{Q}_{2k} + \hat{Q}_{2k+1}$,
where $k$ is an integer corresponding to the spatial site.
Of particular interest are connected correlation functions between spatial charges,\footnote{For periodic boundary conditions, $\langle\hat{ \overline{Q}}_k\rangle  = 0$, but for OBCs $\langle\hat{ \overline{Q}}_k\rangle$ decays exponentially away from the boundaries; see App.~\ref{app:LinvScaling}.} defined as
\begin{equation}
\langle \hat{\overline{Q}}_{j} \hat{\overline{Q}}_{k} \rangle_c \ = \ \langle \hat{\overline{Q}}_{j} \hat{\overline{Q}}_{k} \rangle \ - \
\langle \hat{\overline{Q}}_{j}\rangle  \langle\hat{\overline{Q}}_{k} \rangle \ .
\end{equation}
These correlations decay exponentially for $\lvert j-k\rvert \gtrsim \xi$ due to confinement and charge screening.
Unlike the chiral condensate, which is a sum of single qubit observables, $\langle \hat{\overline{Q}}_{j} \hat{\overline{Q}}_{k} \rangle_c$ is sensitive to correlations between qubits, i.e., requires measurement of $\langle \hat{Z}_j \hat{Z}_k \rangle$.
The results from {\tt ibm\_cusco}  for $L=50$ are shown in Fig.~\ref{fig:qiqj_40}.
The correlations are symmetric under $j\leftrightarrow k$, 
and only the lower-triangle of the correlation matrix is shown.
Each measured value is within $3\sigma$ of the MPS simulator result, consistent with statistical fluctuations.
Also shown in Fig.~\ref{fig:qiqj_40} are the spatial charge-charge correlations as a function of distance, averaged over the lattice volume,
\begin{equation}
\langle \hat{\overline{Q}} \hat{\overline{Q}} \rangle_c (d) \ = \ \frac{1}{L-4-d}\sum_{k=2}^{L-3-d} \langle \hat{\overline{Q}}_k \hat{\overline{Q}}_{k+d} \rangle_c \ .
\end{equation}
To reduce the effects of the boundaries, this sum omits the first and last two spatial lattice sites.
As anticipated, this correlation function decays exponentially, with a characteristic length scale proportional to $\xi = 1/m_{\text{hadron}}$.\footnote{The exponential decay of charge-charge correlations provides motivation for the construction of a truncated Hamiltonian where charge-charge terms are omitted beyond a certain distance. See App.~\ref{app:QoSM} for more details.}
For $d>2$, the correlations are consistent with zero within $2\sigma$ (note that the log scale distorts the error bars),
and 
increased numbers of shots and twirlings 
are needed to distinguish additional points from zero.
The local chiral condensate and charge-charge correlations corresponding to the other values of $L$ are given in App.~\ref{app:qusimDetail}.

\section{Summary and Outlook}
\label{sec:Conclusions}
\noindent
In this work, the vacuum of the lattice Schwinger model was prepared 
on up to 100 qubits of IBM's 
127-qubit {\tt Eagle}-processor quantum computers,
{\tt ibm\_brisbane} and {\tt ibm\_cusco}.
This was accomplished with SC-ADAPT-VQE, an algorithm for identifying systematically improvable
state preparation quantum circuits that can be robustly scaled to operate on any number of qubits
The utility of scalable circuits relies on physically relevant systems often having a (discrete) translational symmetry, and a finite correlation length set by the mass gap.
Together, these imply that the state preparation circuits have unique structure over approximately a correlation length~\cite{Klco:2019yrb,Klco:2020aud},
which is replicated across the lattice.
The lattice Schwinger model with OBCs was chosen to explore these ideas as its vacuum has (approximate) translational invariance and, due to confinement, has a mass gap.
By performing SC-ADAPT-VQE on a classical simulator, state preparation circuits for lattices of $L\leq 14$ (28 qubits) were built from an operator pool containing both translationally invariant terms and ones localized to the boundaries.
Exponential convergence in the quality of the prepared state with both system size and circuit depth enabled the extrapolation of circuits that can be scaled to arbitrarily large lattices.
This methodology was successfully demonstrated by preparing the Schwinger model vacuum on up to 100 superconducting qubits of IBM's quantum computers.
Both the charge-charge correlators and the chiral condensate were measured, and were found to agree with results from an MPS simulator, within statistical uncertainty.
Vital to the success of these quantum simulations involving a large number of qubits was the development of an improved error mitigation technique, which we have called Operator Decoherence Renormalization (ODR).

Due to its generality, 
we expect that the scalable circuit framework embodied by SC-ADAPT-VQE can be applied to other gapped theories with translationally-invariant ground states.
Of particular importance is QCD,  
for which the initialization of ground states for quantum simulations continues to be a daunting prospect.
It is likely that many of the ideas used to construct efficient state preparation circuits for the Schwinger model can be applied to the initialization of the ground state of QCD.
Of course, the operator pool that informs the state preparation circuits will be  
more diverse since the  gauge field is no longer completely constrained by Gauss's law.
Local quark-field operators, extended quark operators with associated gauge links, and closed loops of gauge links will need to be included in the pool.
It is also expected that the variational parameters defining the 
ground-state preparation circuits will converge exponentially, 
once the simulation volume can completely contain the pion(s).

The utility of SC-ADAPT-VQE is that it provides a straightforward prescription for determining low-depth quantum circuits that prepare the ground state on systems of arbitrary size with only classical computing overhead.
This not only allows for the quantum simulation of ground state properties, but will be important for future simulations of dynamics, where preparing the initial state is a crucial first step.
Scalable circuits can
likely be used to prepare single- and multi-hadron states,
for example, a vector meson in the Schwinger model or a baryon in QCD.
Once these states are initialized, they can be used to simulate scattering, 
electroweak processes
and probe properties of dense matter. 
As an example, localized $e^+e^-$ pairs on top of the Schwinger model vacuum could be prepared and evolved 
forward in time. 
Such localized distributions can have high energy components, 
whose propagation through the lattice 
leaves behind showers of particles.
These processes probe the dynamics of 
fragmentation, confinement, 
and hadron production, and
lead to long-range correlations that entangle distant regions of the lattice (see, e.g., Refs.~\cite{Verdel:2019chj,Milsted:2020jmf,Florio:2023dke,Belyansky:2023rgh}).
As simulations of highly-entangling dynamics at scale are beyond the capabilities of classical computing, they are a candidate for an early quantum advantage for scientific applications.

\begin{acknowledgements}
\noindent
We thank Caroline Robin for interesting discussions and reading the manuscript,
and Sophia Economou for interesting discussions about ADAPT-VQE and symmetries.
We also thank the participants of the {\it Quantum Error Mitigation for Particle and Nuclear Physics} IQuS-INT workshop held between May 9-13, 2022 for helpful discussions.
Roland Farrell thanks the organizers of the Quantum Connections summer school where some of this work was carried out.
This work was supported, in part, by 
the U.S. Department of Energy grant DE-FG02-97ER-41014 (Farrell), 
by U.S. Department of Energy, Office of Science, Office of Nuclear Physics, Inqubator for Quantum Simulation (IQuS)\footnote{\url{https://iqus.uw.edu/}}
under Award Number DOE (NP) Award DE-SC0020970 
via the program on Quantum Horizons: QIS Research and Innovation for Nuclear Science\footnote{\url{https://science.osti.gov/np/Research/Quantum-Information-Science}}
(Ciavarella,  Farrell,  Savage), 
and the 
Quantum Science Center (QSC),\footnote{\url{https://qscience.org}} 
a National Quantum Information Science Research Center of the U.S.\ Department of Energy (DOE) (Illa).
This work was also supported, in part, through the Department of Physics\footnote{\url{https://phys.washington.edu}} 
and the College of Arts and Sciences\footnote{\url{https://www.artsci.washington.edu}}
at the University of Washington.
This research used resources of the Oak Ridge Leadership Computing Facility, 
which is a DOE Office of Science User Facility supported under Contract DE-AC05-00OR22725.
We acknowledge the use of IBM Quantum services for this work. The views expressed are those of the authors, and do not reflect the official policy or position of IBM or the IBM Quantum team. This work was enabled, in part, by the use of advanced computational, storage and networking infrastructure provided by the Hyak supercomputer system at the University of Washington.\footnote{\url{https://itconnect.uw.edu/research/hpc}}
We have made extensive use of Wolfram {\tt Mathematica}~\cite{Mathematica},
{\tt python}~\cite{python3,Hunter:2007}, {\tt julia}~\cite{Julia-2017},
{\tt jupyter} notebooks~\cite{PER-GRA:2007} 
in the {\tt Conda} environment~\cite{anaconda},
and IBM's quantum programming environment {\tt qiskit}~\cite{qiskit}. The DMRG calculations were performed using the C++ {\tt iTensor} library~\cite{fishman2022itensor}.
\end{acknowledgements}

\clearpage
\appendix

\section{The Schwinger Model Hamiltonian in the \texorpdfstring{$Q=0$}{Q=0} Sector}
\label{app:QoSM}
\noindent
In the absence of background electric charges, 
the lowest energy sector of the Schwinger model has vanishing charge,
$Q=\sum_n Q_n  =0$.
Restriction to this sector permits a simplification of the Hamiltonian,
reducing the number of terms appearing in the electric contribution:
\begin{align}
\hat H_{el}
    & \ \stackrel{Q=0}{=} \ \frac{g^2}{2}\left \{ \sum_{n=0}^{L-1}\left (L-n\right )\left[\hat Q_n^2 + (1-\delta_{n0})\hat Q_{2L-n}^2 \right] \ + \ 2\sum_{n=0}^{L-1}\left [\sum_{m=n+1}^{L-1}\left (L-m \right )\hat{Q}_m \hat{Q}_n \ + \ \sum_{m=1}^{n-1}(L-n)\hat{Q}_{2L-m}\hat{Q}_{2L-n}\right ]
\right  \} \ .
\end{align}
This reduces the number of $\hat Q_n \hat Q_m$ terms from $1-3L+2L^2$ to $1+L^2-2L$ and the required connectivity from all-to-all to half-to-half. 
Note that this can also be used to simplify the $1+1$D $SU(N)$ Hamiltonian in the $Q_n^{(a)} = 0$ (color singlet) sector, 
by replacing $\hat Q_n \to \hat Q_n^{(a)}$ with $a\in\{1,2,\ldots, N^2-1\}$.

Further simplifications to this Hamiltonian are likely possible by taking advantage of the exponential 
decay of correlations between spatial charges $\hat{\overline{Q}}$ separated by distance $d>\xi$.
This will allow for the construction of a truncated $\hat{H}_{el}$ that only has ${\mathcal O}(L \xi)$ terms.
In addition, such a Hamiltonian will only require connectivity between qubits separated by $d\lesssim\xi$ instead of $d\le L$.

\clearpage
\section{Volume Extrapolation of the Energy Density and Chiral Condensate}
\label{app:LinvScaling}
\begin{figure}[ht!]
    \centering
    \includegraphics[width=0.7\columnwidth]{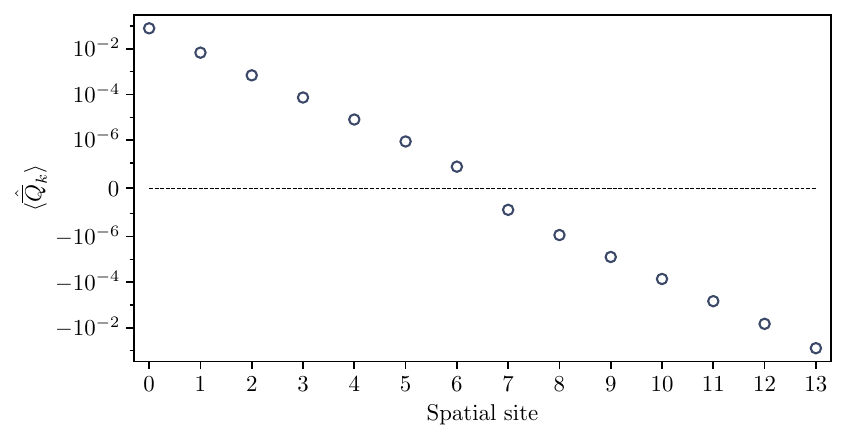}
    \caption{Charge on each spatial site, $\hat{\overline{Q}}_k$, for $m=0.5,g=0.3$ and $L=14$ obtained from exact diagonalization of the Hamiltonian. 
}
    \label{fig:Qboundary}
\end{figure}

\noindent
Here the vacuum energy density and chiral condensate are extrapolated to $L=\infty$.
The results of exact diagonalization and DMRG calculations are considered independently, providing consistent results within uncertainties. For the DMRG calculations, 60 sweeps were performed with a maximum allowed bond dimension of 150 and a truncation of Schmidt coefficients below $10^{-10}$. This showed a convergence of $10^{-10}$ in the energy of the vacuum state.
Discussions in Sec.~\ref{sec:infVolExtra} motivated an inverse-power, $1/L$, dependence 
of the exact vacuum energies as the infinite-volume limit is approached. 
This scaling was argued when $L$ is much larger than the longest correlation length, and with OBCs.
Therefore, for masses and couplings that give rise to the lowest-lying hadron 
being completely contained within the lattice volume,
we anticipate functional forms
\begin{align}
 \varepsilon(L) & = \varepsilon(\infty)\ +\ \frac{e_1}{L}\ +\ {\cal O}\left(\frac{1}{L^2}\right) \ ,\qquad\ 
 \chi(L)  = \chi(\infty)\ +\ \frac{d_1}{L}\ +\ {\cal O}\left(\frac{1}{ L^2}\right)
 \ ,
\end{align}
for $\varepsilon$ and $\chi$.
This is due to the finite penetration depth of boundary effects, and the exponential convergence of both the volume and the surface contributions to their infinite-volume values.
As a result, the surface terms make 
${\cal O}\left(1/L\right)$ contributions to intensive quantities, e.g., densities.
To illustrate this, the expectation value of the charge on each spatial site, $\hat{\overline{Q}}_k$, for $m=0.5,g=0.3$ and $L=14$ is shown in Fig.~\ref{fig:Qboundary}.
This converges exponentially with the distance to the boundary to $\langle \hat{\overline{Q}}_k \rangle = 0$, the expected infinite volume value.

\begin{figure}[ht!]
    \centering
    \includegraphics[width=\columnwidth]{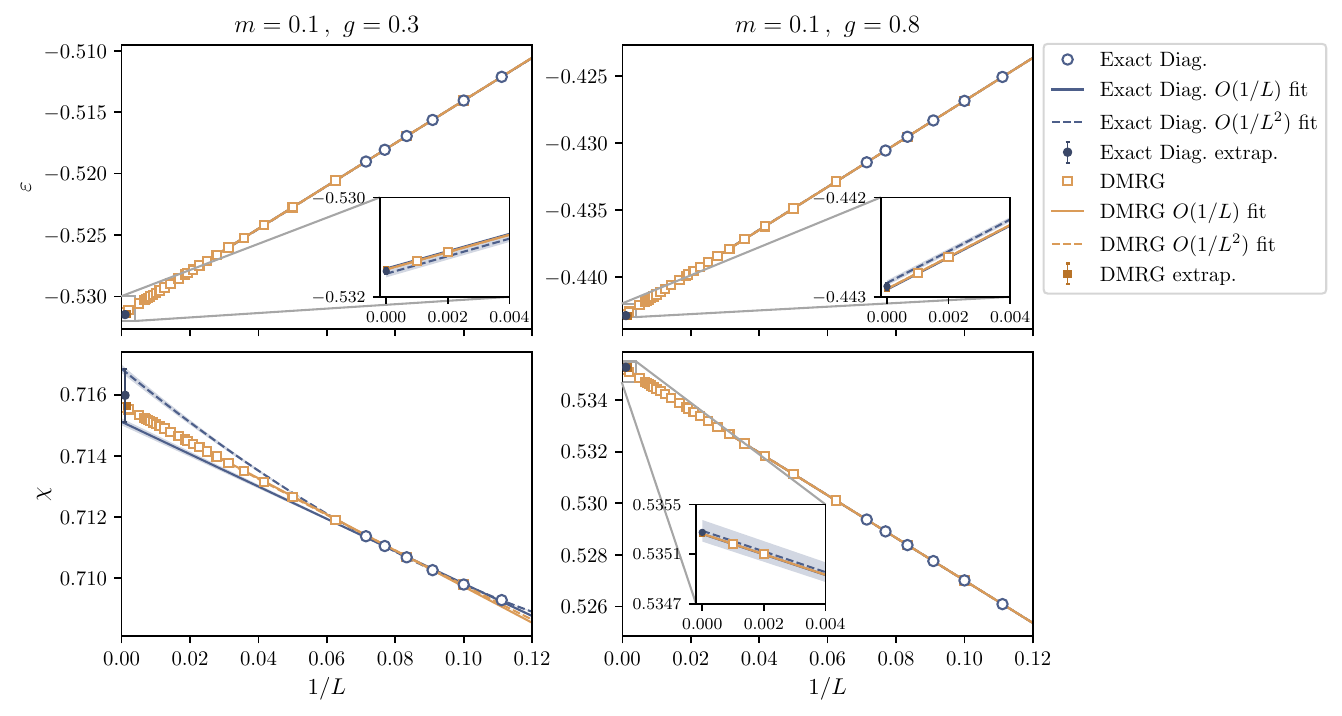}
    \caption{$L$-extrapolations of the vacuum energy density $\varepsilon$ (top) and chiral condensate $\chi$ (bottom) for $m=0.1$ and $g=0.3$ (left) and $m=0.1$ and $g=0.8$ (right).
    Each panel shows extrapolations of the exact values given in Tables~\ref{tab:Echifmp1gp3} and~\ref{tab:Echifmp1gp8} (blue circles) and of the results from DMRG calculations (orange squares) given in Table~\ref{tab:DMRGresults} for $L\ge 9$.
    The solid lines correspond to linear extrapolations and the dashed lines correspond to quadratic extrapolations, and are found to overlap (see the insets).
    The darker points show the $L=\infty$ extrapolated value, with an uncertainty determined by the difference between the linear and quadratic extrapolations.
}
    \label{fig:EdensityLextrap}
\end{figure}
The results of fits to the exact and DMRG results for the energy density and chiral condensate for $m=0.5, g=0.3$ are shown in Fig.~\ref{fig:EdChiLextrap53}, 
and for $m=0.1, g=0.8$ and $m=0.1, g=0.3$ are shown in Fig.~\ref{fig:EdensityLextrap}.
Using polynomials 
that are linear and quadratic in $1/L$,
fits  are performed for $L\geq 9$ and extrapolated to $L=\infty$.
The differences between extrapolations obtained from the two fit forms are used to 
estimate the systematic fitting error, 
corresponding to the black and grey points (and error bars).
The difference between linear and quadratic fits is negligible for the exact results, except 
for the chiral condensate 
in the case of $m=0.1$ and $g=0.3$, which sees a small quadratic dependence.  
When the fit interval is reduced to $L\ge 10$, this dependence once again becomes negligible.

\begin{table}[!ht]
\renewcommand{\arraystretch}{1.4}
\begin{tabularx}{\textwidth}{|c || Y | Y || Y | Y || Y | Y |}
\hline
 & \multicolumn{2}{c||}{$m=0.1, g=0.3$} & \multicolumn{2}{c||}{$m=0.1, g=0.8$} & \multicolumn{2}{c|}{$m=0.5, g=0.3$} \\ \hline
 $L$ & $\varepsilon^{(\text{DMRG})}$ & $\chi^{(\text{DMRG})}$ & $\varepsilon^{(\text{DMRG})}$ & $\chi^{(\text{DMRG})}$ & $\varepsilon^{(\text{DMRG})}$ & $\chi^{(\text{DMRG})}$ \\
 \hline
 \hline
10   & -0.51405 & 0.70979 & -0.42685 & 0.52701 & -0.31707 & 0.33358 \\ \hline
12   & -0.51694 & 0.71068 & -0.42953 & 0.52838 & -0.31936 & 0.33517 \\ \hline
16   & -0.52057 & 0.71190 & -0.43288 & 0.53010 & -0.32221 & 0.33716 \\ \hline
20   & -0.52275 & 0.71265 & -0.43488 & 0.53114 & -0.32393 & 0.33836 \\ \hline
24   & -0.52420 & 0.71315 & -0.43622 & 0.53182 & -0.32508 & 0.33916 \\ \hline
28   & -0.52523 & 0.71350 & -0.43718 & 0.53232 & -0.32589 & 0.33973 \\ \hline
32   & -0.52601 & 0.71377 & -0.43790 & 0.53268 & -0.32650 & 0.34016 \\ \hline
36   & -0.52661 & 0.71398 & -0.43846 & 0.53297 & -0.32698 & 0.34049 \\ \hline
40   & -0.52710 & 0.71414 & -0.43890 & 0.53320 & -0.32736 & 0.34075 \\ \hline
44   & -0.52749 & 0.71428 & -0.43927 & 0.53339 & -0.32768 & 0.34097 \\ \hline
48   & -0.52782 & 0.71439 & -0.43957 & 0.53354 & -0.32794 & 0.34115 \\ \hline
52   & -0.52810 & 0.71449 & -0.43983 & 0.53368 & -0.32816 & 0.34130 \\ \hline
54   & -0.52823 & 0.71453 & -0.43994 & 0.53374 & -0.32825 & 0.34137 \\ \hline
60   & -0.52855 & 0.71464 & -0.44024 & 0.53389 & -0.32851 & 0.34155 \\ \hline
70   & -0.52896 & 0.71479 & -0.44062 & 0.53408 & -0.32883 & 0.34178 \\ \hline
80   & -0.52927 & 0.71489 & -0.44091 & 0.53423 & -0.32908 & 0.34195 \\ \hline
90   & -0.52952 & 0.71498 & -0.44113 & 0.53435 & -0.32927 & 0.34208 \\ \hline
100  & -0.52971 & 0.71504 & -0.44131 & 0.53444 & -0.32942 & 0.34219 \\ \hline
110  & -0.52987 & 0.71510 & -0.44146 & 0.53451 & -0.32955 & 0.34228 \\ \hline
120  & -0.53000 & 0.71514 & -0.44158 & 0.53458 & -0.32965 & 0.34235 \\ \hline
130  & -0.53011 & 0.71518 & -0.44168 & 0.53463 & -0.32974 & 0.34241 \\ \hline
140  & -0.53021 & 0.71521 & -0.44177 & 0.53467 & -0.32981 & 0.34246 \\ \hline
150  & -0.53029 & 0.71524 & -0.44185 & 0.53471 & -0.32988 & 0.34251 \\ \hline
200  & -0.53058 & 0.71534 & -0.44212 & 0.53485 & -0.33011 & 0.34267 \\ \hline
500  & -0.53110 & 0.71552 & -0.44260 & 0.53510 & -0.33052 & 0.34295 \\ \hline
1000 & -0.53128 & 0.71558 & -0.44276 & 0.53518 & -0.33066 & 0.34305 \\ \hline
\end{tabularx}
\caption{Additional results for the energy density $\varepsilon$ and chiral condensate $\chi$ obtained from DMRG calculations, and used in the extrapolations in Figs.~\ref{fig:EdChiLextrap53} and~\ref{fig:EdensityLextrap}.
}
 \label{tab:DMRGresults}
\end{table}

\clearpage
\section{Optimizing Trotterized Circuits for State Preparation}
\label{app:TrotterErrors}
\noindent
As discussed in the main text, even after the operator pool has been chosen for SC-ADAPT-VQE, there remains freedom in how the pool of unitary operators, Eq.~(\ref{eq:TrotOp}), 
is implemented as quantum circuits.
For example, 
instead of leading-order Trotterization,
a higher-order Trotterization could be used to suppress Trotter errors. 
Alternatively, 
different orderings of the terms in the leading-order Trotterization can be considered.
This freedom can be used to optimize the convergence of SC-ADAPT-VQE with circuit depth.
Also, different Trotter orderings can break the CP symmetry.
The circuit orderings in Fig.~\ref{fig:ohd35multicirc} were chosen to minimize the circuit depth, and for $d=1,3,5$ this ordering preserves CP, while for $d=7,9$ it breaks CP.

\begin{figure}[ht!]
    \centering
    \includegraphics[width=\columnwidth]{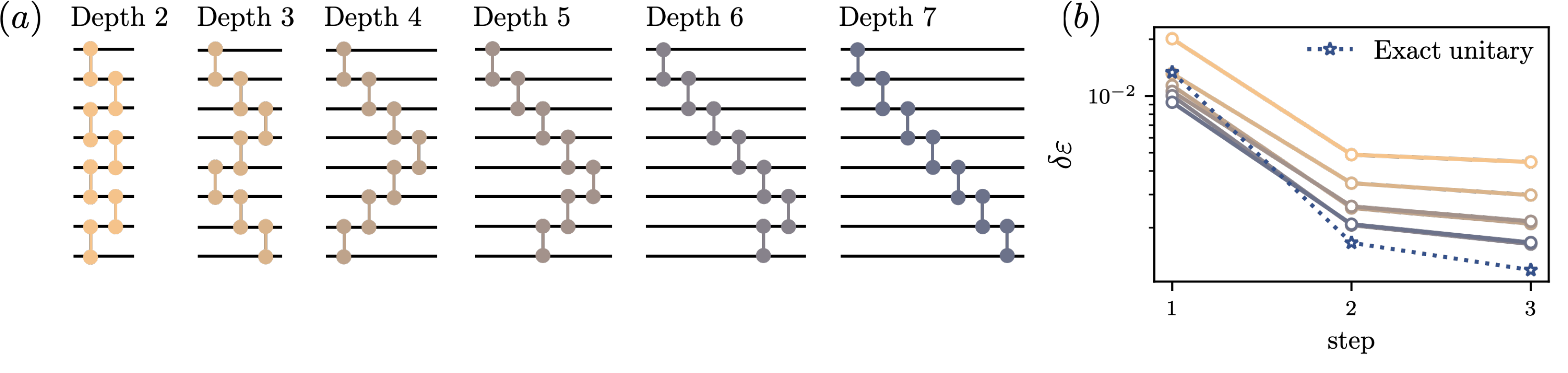}
    \caption{(a) Orderings of Trotterized terms 
    for implementing $\hat{O}^{V}_{mh}(1)$. 
    Circuits of depth 2 to 7 are shown from left to right, with the dumbbells representing the circuit in Fig.~\ref{fig:RpmOh35circ}a.
    (b) Deviations in the energy density of the SC-ADAPT-VQE prepared state for $m=0.5, g=0.3$ and $L=10$
    with different depth implementations of the Trotterization of $\hat{O}^{V}_{mh}(1)$.
    }
    \label{fig:PyramidTrotter}
\end{figure}

Consider the different arrangements of the terms in the Trotterization of $\hat{O}^{V}_{mh}
(1)$, given in Eq.~(\ref{eq:poolComm}), as shown in Fig.~\ref{fig:PyramidTrotter}a.
The depth-2 ordering (left) was used 
to obtain the results presented 
in the main text as it leads to the shallowest circuits.
All the orderings shown
in Fig.~\ref{fig:PyramidTrotter}a
are equivalent up to $\mathcal{O}[(\theta_1)^2]$ 
(where $\theta_1$ is 
the coefficient of the operator in the corresponding unitary operator),
but the deeper circuits allow for the generation of longer-range correlations. 
Note that the deeper circuits can break the CP symmetry; e.g. for $L=10$ depths 2 and 4 preserve CP while depths 3, 4, 5 and 7 break CP.
It is found that this added circuit depth improves the convergence of SC-ADAPT-VQE, as shown in Fig.~\ref{fig:PyramidTrotter}b.
This demonstrates that to minimize circuit depth, for a fixed error threshold, it is preferable to choose a deeper Trotterization of $\hat{O}^{V}_{mh}(1)$, 
instead of 
going to a greater number of SC-ADAPT-VQE steps.
For example, it is more efficient to perform 2 steps of SC-ADAPT-VQE with a depth-3 Trotterization of $\hat{O}^{V}_{mh}(1)$, than to perform 3 steps of SC-ADAPT-VQE with a depth-2 Trotterization of $\hat{O}^{V}_{mh}(1)$.
Also shown in Fig.~\ref{fig:PyramidTrotter}b are results obtained 
from performing SC-ADAPT-VQE with exact unitary operators (no Trotterization).
This is found to always perform better than the Trotterized unitaries, 
except for a single step.
Intriguingly, for a single step, the error is less with a deep first-order Trotterization than with the exact unitary.
This suggests that the optimizer is finding a solution in which the Trotter errors are tuned to improve the overlap with the vacuum.
Note that the deeper Trotterizations of $\hat{O}^{V}_{mh}(1)$ move the recurrence of $\hat{O}^{V}_{mh}(1)$ (e.g., at step 4 for $m=0.5, g=0.3$) to deeper in the SC-ADAPT-VQE ansatz.

\clearpage
\section{Volume Extrapolations of the SC-ADAPT-VQE Variational Parameters: 
An ``effective-\texorpdfstring{$\theta_i^{\infty}$}{theta infinity}''}
\label{app:ThetaScaling}
\noindent
To initialize large quantum registers, the variational parameters defining the state-preparation quantum circuits need to be extrapolated with high precision. 
In volumes large enough to contain the longest correlation length, 
the variational parameters are expected to be exponentially close to their infinite-volume values.
Therefore, we assume that the form of the volume dependence for practical purposes is that given in Eq.~(\ref{eq:thetaextrap}),
\begin{equation}
\theta_i(L) \ = \ \theta_i^{\infty} \ + \  c_1\,e^{-c_2 \, L} 
\ ,
\label{eq:thetaextrapAPP}
\end{equation}
and check the self-consistency of this form.\footnote{For the current paper, due to the small number of parameters, the selection of the points to be fitted was determined by visual inspection (if the points followed an exponential decay or not).}
While there could be a polynomial coefficient of the exponential, 
we find that this is not required.
Fitting exponential functions can be challenging; 
however, with results over a sufficient range of $L$, algebraic techniques, such as effective masses, have proven useful in lattice QCD calculations to eliminate ``uninteresting'' parameters, while at the same time mitigating correlated fluctuations in measurements~\cite{Michael:1985ne,Luscher:1990ck,DeGrand1990FromAT,Fleming:2004hs,Beane:2009kya}.
With the goal of initializing large lattices, it is the
$\theta_i^{\infty}$ that are of particular interest.

Assuming the volume dependence given in Eq.~(\ref{eq:thetaextrapAPP}),
it is useful to form four relations
\begin{align}
y_L & = \theta_i(L) \ - \ \theta_i^{\infty}\ = \ c_1\,e^{-c_2 \, L} 
\ \ ,\ \ 
y_{L+1}\ =\ \theta_i(L+1) \ - \ \theta_i^{\infty}\ = \ c_1\, e^{-c_2} e^{-c_2 \, L} 
\nonumber \\
y_{L+2} & = \theta_i(L+2) \ - \ \theta_i^{\infty}\ = \ c_1\, e^{-2 c_2} e^{-c_2 \, L} 
\ \ ,\ \ 
y_{L+3}\ =\ \theta_i(L+3) \ - \ \theta_i^{\infty}\ = \ c_1\,e^{-3 c_2} e^{-c_2 \, L} 
\ .
\label{eq:LdifsA}
\end{align}
These relations can be combined to isolate $\theta_i^{\infty}$, providing an $L$-dependent ``effective-$\theta_i^{\infty}$'', denoted as $\theta_{i, {\rm eff}}^{\infty}$:
\begin{align}
& y_{L+1} y_{L+2} \  =\  y_{L}  y_{L+3} \ ,
\nonumber \\
& \theta_{i, {\rm eff}}^{\infty}(L) \  =\  
\frac{\theta_i(L) \theta_i(L+3) - \theta_i(L+1)\theta_i(L+2)}{
\theta_i(L) + \theta_i(L+3) - \theta_i(L+1) - \theta_i(L+2)
}
\ .
\label{eq:LdifsB}
\end{align}
For a sufficiently large set of results, 
$\theta_{i, {\rm eff}}^{\infty}(L)$ 
will plateau for large $L$ if the functional form in 
Eq.~(\ref{eq:thetaextrapAPP}) correctly describes the results.
This plateau can be fit by a constant,
over some range of large $L$,
to provide an estimate of $\theta_i^{\infty}$.
This method is similar to using {\tt varpro} (variable projection) in a multi-parameter 
$\chi^2$-minimization.

\begin{figure}[ht!]
    \centering
    \includegraphics[width=0.9\columnwidth]{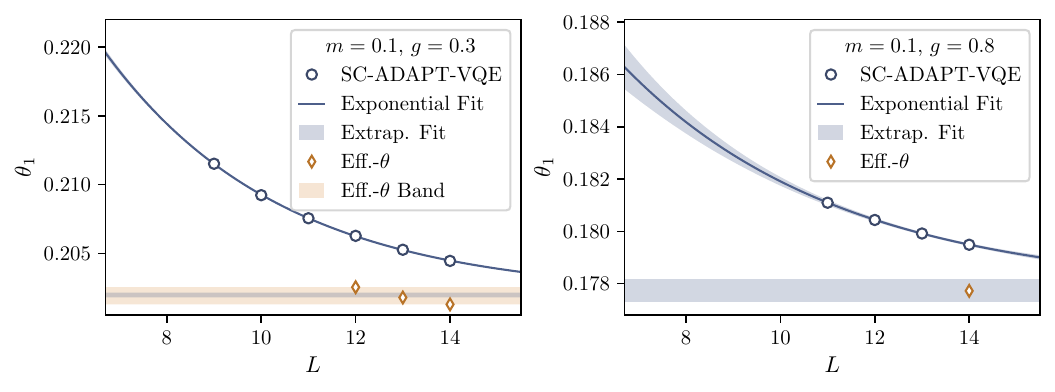}
    \caption{
    Further examples of fitting the asymptotic $L$-dependence of the variational parameters defining the state-preparation quantum circuit, determined from classical simulations using SC-ADAPT-VQE.
    The results for $\theta_1=\hat O_{mh}^{V}(1)$ (blue points) for 
    $m=0.1$ and $g=0.3$ (left panel) given in Table~\ref{tab:AnglesXCircmp1gp3} and
    $m=0.1$ and $g=0.8$ (right panel) given in Table~\ref{tab:AnglesXCircmp1gp8}
    are extrapolated to $L=\infty$ by 
i) use of a 3-parameter fit given in Eq.~(\ref{eq:thetaextrap}), as shown by the blue line and shaded region, 
with an asymptotic value shown by the gray region,
and by
ii) forming of effective-$\theta$ (orange diamonds) 
with the maximum and minimum values shown as the orange shaded region (where possible).
}
    \label{fig:ThetaLextrap}
\end{figure}

As an example, the results for $\theta_1^{\infty}$ 
from a 3-parameter fit of $\theta_1$ to Eq.~\eqref{eq:thetaextrapAPP} are compared with a determination using
$\theta_{1, {\rm eff}}^{\infty}(L)$  from
Eq.~(\ref{eq:LdifsB}).
Results obtained with these two methods for $m=0.1, g=0.3$ and for $m=0.1, g=0.8$ are shown in Fig.~\ref{fig:ThetaLextrap}.
The result obtained 
from fitting a constant to $\theta_{1, {\rm eff}}^{\infty}(L)$ 
is consistent with the asymptotic result from the 3-parameter fit, but with somewhat larger uncertainty.
The current deficiency of this comparison is the small number of points in the plateau region, and results for larger $L$ are required for a more complete comparison.
Analysis of the other variational parameters shows a similar behavior.

The consistency between the two extraction methods is encouraging, and suggests that the selected exponential form may indeed well describe the results.
The fitting method is likely insensitive to polynomial corrections (coefficients), and requires further exploration to fully-quantify uncertainties in these asymptotic values of the variational parameters.
However, as the MPS simulations with these extrapolated angles reproduce the results calculated with DMRG, it appears that, for the systems and parameters we have selected in our analysis, systematic errors introduced by selecting this functional form are small.

\clearpage
\section{Operator Decoherence Renormalization (ODR)}
\label{app:QSimError}
\noindent
To mitigate the effects of noise, the decoherence renormalization technique~\cite{Urbanek_2021,ARahman:2022tkr,Farrell:2022wyt,Ciavarella:2023mfc} 
is modified for use with larger systems. 
In its original form, 
decoherence renormalization assumes that each qubit decoheres at the same rate under a depolarizing noise channel.
When working with a small number of qubits, this is a reasonable approximation, but for larger systems, it is necessary to consider the 
rate of decoherence of each qubit individually.
After Pauli twirling, the qubit errors are well described by a Pauli error channel~\cite{PhysRevA.72.052326}, which maps the $N$ qubit density matrix to
\begin{equation}
\rho \ \to  \ \sum_{i=1}^{4^N} \eta_i \hat{P}_i \rho \hat{P}_i \ , 
\end{equation} 
where $\hat{P}_i$ is a tensor product of Pauli operators ($\hat{I}$, $\hat{X}$, $\hat{Y}$ or $\hat{Z}$) acting on $N$ qubits, and the set of $\eta_i$ characterizes the error channel. It is important to understand the effect of this error channel on observables. Generic observables can be written as a sum over tensor products of Pauli operators, so it suffices to consider an observable, $\hat{O}$, that is a tensor product of Pauli operators. Under a Pauli error channel, the measured (noisy) expectation value, $\langle \hat{O}\rangle_{\text{meas}}$, is given by
\begin{equation}
\langle \hat{O} \rangle_{\text{meas}} \ =  \sum_{i=1}^{4^N} \eta_i\Tr( \hat{P}_i \hat{O} \hat{P}_i \rho) \ .
\end{equation}
Note that $\hat{P}_i \hat{O} \hat{P}_i = \pm \hat{O}$, depending on whether or not $\hat{O}$ and $\hat{P}_i$ commute or anti-commute. Using this fact, the measured (noisy) expectation value $\langle \hat{O}\rangle_{\text{meas}}$, can be seen to be directly proportional to the predicted (noiseless) expectation value, $\langle \hat{O}\rangle_{\text{pred}} = {\Tr}(\hat{O} \rho)$, i.e.,
\begin{equation}
\langle \hat{O} \rangle_{\text{meas}} \ =  \left ( 1-\eta_O\right ) \langle \hat{O }\rangle_{\text{pred}} \ .
\label{eq:etaO}
\end{equation}

The ODR factor $\eta_O$ is, in general, 
distinct for each operator, and can be estimated by running a mitigation circuit that has the same structure as the physics circuit, but where $\langle \hat{O} \rangle_{\text{pred}}$ is already known.
In this work, the mitigation circuit was taken to be the state preparation circuit with variational parameters set to zero, which is the identity in the absence of noise.
This mitigation circuit will have the same noise channel as the physics circuit provided that the noise is dominated by errors in the two-qubit gates and is independent of the single qubit rotation angles in the circuit. Without noise, the mitigation circuit prepares the strong coupling vacuum, where  $\langle \hat{O} \rangle_{\text{pred}}$ is known, 
and therefore $\eta_O$ can be computed.
Once $\eta_O$ is determined, Eq.~\eqref{eq:etaO} is used to estimate the  value of the noiseless observable from the 
results of the physics circuits.

An added benefit of ODR is that it reduces the need for other error mitigation techniques.
For example, readout errors are 
partially mitigated since the measured observables are affected by both gate and measurement errors.
This is convenient as current measurement mitigation techniques require a large classical computing overhead.
It also reduces the need for post-selection, which in our work could have been performed on states with total charge $Q = 0$.
This post-selection removes single-qubit errors, but introduces further correlations between qubits. These correlations effectively increase the size of the 
single-qubit errors (making observables sensitive to errors anywhere on the register).
This reduces the efficacy of the Pauli error model, making post-selection incompatible with ODR.\footnote{This is not true for observables involving the entire qubit register, 
e.g., the vacuum-vacuum persistence probability. 
This is because applying the $Q=0$ constraint when measuring global observables will not introduce any new correlations.}
Another desirable feature of ODR is that it allows simulations to retain the results of 
a much larger fraction of the ensemble. 
This is because 
the probability of a single-qubit error increases with system size, 
and therefore much of the ensemble is lost with naive post selection.
Further, such errors have little effect on local observables that are summed across the entire qubit register.

\clearpage
\section{Additional Results From Classical Simulations}
\label{app:classData}
\noindent
The results corresponding to Fig.~\ref{fig:ClassConvXCirc} for $m=0.1,g=0.3$ are given in Table~\ref{tab:AnglesXCircmp1gp3} and Table~\ref{tab:Echifmp1gp3}, and for
$m=0.1,g=0.8$ are given in Table~\ref{tab:AnglesXCircmp1gp8} and Table~\ref{tab:Echifmp1gp8}.
The $6^{\rm th}$ step of the algorithm is chosen for $m=0.1, g=0.8$ because the operator structure through $L=14$ has converged, allowing a consistent extrapolation of the circuits to large $L$.
This can be seen by comparing the operator structure in Table~\ref{tab:AnglesXCircmp1gp8} (6 steps) and Table~\ref{tab:AnglesXCircmp1gp8Step7} (7 steps). An interesting observation is that the sum of parameters for a particular operator in the ansatz remains approximately unchanged when an additional insertion of the operator is added. 
For example, compare the sum of parameters for $\hat{O}^V_{mh}(1)$ between $L=8$ and $9$ in Table~\ref{tab:AnglesXCircmp1gp3}.
Using the same method as for $m=0.5, g=0.3$ in Sec.~\ref{sec:ClassSim}, scalable circuits for $m=0.1, g=0.3$ and $m=0.1$ and $g=0.8$ were also determined.
The results of running these circuits on {\tt qiskit}'s MPS simulator for $m=0.1, g=0.3$ and $m=0.1$ and $g=0.8$ are given in Table~\ref{tab:SQCResultsApp}.
Due to the longer correlation lengths for these parameters, it was not possible to go to $L=500$ with the available computing resources.
In these MPS simulations, {\tt qiskit}'s default settings were used, where the bond dimension increases until machine precision is achieved.
The details of the qiskit MPS simulator can be found on the qiskit website~\cite{qiskitMPS}.
Again, the energy density and chiral condensate are found to have precision comparable to 
that found on smaller systems.
This shows that, despite the longer correlation lengths for $m=0.1, g=0.3$ and $m=0.1,g=0.8$, it is still possible to accurately extrapolate the state preparation circuits to large lattices. 
Note that stabilization of operator ordering for the different $m$ and $g$ (see Tables \ref{tab:AnglesXCircmp5gp3}, \ref{tab:AnglesXCircmp1gp3} and \ref{tab:AnglesXCircmp1gp8Step7}) does not follow the hierarchy in correlation lengths. 
This is because larger $\xi$ increases both the contribution of the volume $\sim e^{-d/\xi}$ and surface $\sim \xi/L$ terms to the energy density.

To emphasize the advantage of performing SC-ADAPT-VQE using a classical simulator, we give an estimate of the number of shots required to perform SC-ADAPT-VQE on a quantum computer.
For $m=0.5,g=0.3, L=14$ performing 10 steps of SC-ADAPT-VQE required $\sim 6000$ calls to the optimizer, in addition to about $500$ evaluations of $\langle [\hat{H}, \hat{O}_i ] \rangle$ for pool operators $\hat{O}_i$.
Each one of these calls required roughly $10^{-3}$ precision in the measured observable, corresponding to about $10^6$ shots on a noiseless device.
Therefore, SC-ADAPT-VQE for $L=14$ would require $\sim 10^{10}$ shots on a noiseless device. 
Factoring in the effects of device noise would increase this estimate by at least a factor of $10$, and probably close to a trillion shots would be required to perform SC-ADAPT-VQE on a quantum computer. This is infeasible on current hardware.
\begin{table}[!ht]
\renewcommand{\arraystretch}{1.4}
\begin{tabularx}{\textwidth}{|c || Y | Y | Y | Y | Y ||  Y| Y | Y | Y |}
 \hline
 \diagbox[height=23pt]{$L$}{$\theta_i$} & $\hat O_{mh}^{V}(1)$
 & $\hat O_{mh}^{V}(3)$ & $\hat O_{mh}^{V}(5)$ & $\hat O_{mh}^{V}(1)$ & $\hat O_{mh}^{V}(7)$ &
 $\hat O_{mh}^{V}(9)$  &
 $\hat O_{mh}^{S}(0,1)$ & 
 $\hat O_{mh}^{V}(9)$  &
 $\hat O_{mh}^{V}(1)$  \\
 \hline\hline
 6 & 0.25704 & -0.11697 & 0.04896 & 0.18116 & -0.02664 & -- & 0.19193 & 0.01539 & --\\
 \hline
 7 & 0.25796 & -0.11580 & 0.04776 & 0.18099 & -0.02471 & 0.01250 & 0.18971 & -- & --\\
 \hline
 8 & 0.25862 & -0.11507 & 0.04711 & 0.18087 & -0.02380 & 0.01155 & 0.18832 & -- & --\\
 \hline
 9 & 0.21152 & -0.11560 & 0.04859 & 0.12687 & -0.02419 & 0.01162 & -- & -- & 0.11093\\
 \hline
 10 & 0.20923 & -0.11491 & 0.04809 & 0.12749 & -0.02368 & 0.01122 & -- & -- & 0.11182\\
 \hline
 11 & 0.20755 & -0.11437 & 0.04771 & 0.12792 & -0.02331 & 0.01093 & -- & -- & 0.11244\\
 \hline
 12 & 0.20628 & -0.11393 & 0.04741 & 0.12823 & -0.02303 & 0.01072 & -- & -- & 0.11289\\
 \hline
  13 &0.20526& -0.11357&0.04716&0.12846& -0.02280&0.01056&--&--&0.11324 \\
  \hline
    14 &0.20445& -0.11328&0.04696&0.12863& -0.02262&0.01044&--&--&0.11352\\
 \hline
 \hline
  $\infty$ & 0.202& -0.112&0.046&0.129& -0.022&0.010 & -- & -- & 0.114  \\
 \hline
\end{tabularx}
\caption{
Same as Table~\ref{tab:AnglesXCircmp5gp3} except for $m=0.1, g=0.3$.
}
 \label{tab:AnglesXCircmp1gp3}
\end{table}
\begin{table}[!ht]
\renewcommand{\arraystretch}{1.4}
\begin{tabularx}{\textwidth}{|c || Y | Y || Y |  Y || Y || c|}
 \hline
 $L$ 
 & $\varepsilon^{\rm (aVQE)}$ 
 & $\varepsilon^{\rm (exact)}$ 
 & $\chi^{\rm (aVQE)}$
  & $\chi^{\rm (exact)}$
 & $\infiL$  & \# CNOTS/\text{qubit}
  \\
 \hline\hline
6 & -0.49927& -0.50256 & 0.68192& 0.70834& 0.00377 & 38\\
\hline
7 & -0.50350& -0.50663  & 0.68442& 0.70837 & 0.00337 & 44.3\\
\hline
8 & -0.50670& -0.50971  & 0.68653& 0.70877 & 0.00309 & 48.8\\
\hline
9 & -0.50838& -0.51212  & 0.68694& 0.70928 & 0.00449 & 53.7\\
\hline
10 & -0.51057& -0.51405  & 0.68902& 0.70979 & 0.00412 & 56.1\\
\hline
11 & -0.51236& -0.51563  & 0.69073& 0.71026 & 0.00382 & 58.1\\
\hline
12 & -0.51385& -0.51694 & 0.69217& 0.71068 & 0.00358 & 59.8\\
\hline
13 & -0.51512& -0.51806 & 0.69340& 0.71105  & 0.00337 & 61.2\\
\hline
14 & -0.51620& -0.51902 &0.69445& 0.71137 & 0.00319 & 62.4\\
 \hline
\end{tabularx}
\caption{
Same as Table~\ref{tab:Echifmp5gp3} except for $m=0.1, g=0.3$.
}
 \label{tab:Echifmp1gp3}
\end{table}
\begin{table}[!ht]
\renewcommand{\arraystretch}{1.4}
\begin{tabularx}{\textwidth}{|c || Y | Y | Y | Y | Y ||  Y| Y |}
 \hline
 \diagbox[height=23pt]{$L$}{$\theta_i$} & $\hat O_{mh}^{V}(1)$
 & $\hat O_{mh}^{V}(3)$ & $\hat O_{mh}^{V}(5)$ & $\hat O_{mh}^{V}(1)$ & $\hat O_{mh}^{V}(7)$ &  $\hat O_{mh}^{S}(0,1)$ & 
 $\hat O_{mh}^{V}(1)$ \\
 \hline\hline
 6 & 0.22698 & -0.06357 & 0.01441 & 0.15594 & -0.00418 & 0.14247 & --\\
 \hline
 7 & 0.22784 & -0.06303 & 0.01416 & 0.15559 & -0.00395 & 0.14111  & --\\
 \hline
 8 & 0.22843 & -0.06267 & 0.01401 & 0.15535 & -0.00382 & 0.14018 & -- \\
 \hline
 9 & 0.22885 & -0.06240 & 0.01390 & 0.15518 & -0.00374 & 0.13951 & --\\
 \hline
 10 & 0.22918 & -0.06219 & 0.01382 & 0.15505 & -0.00368 & 0.13900 & --\\
 \hline
 11 & 0.18110 & -0.06192 & 0.01431 & 0.11095 & -0.00377 & -- & 0.09796 \\
 \hline
 12 & 0.18044 & -0.06169 & 0.01423 & 0.11108 & -0.00372 & -- & 0.09809 \\
 \hline
 13 &0.17992 & -0.06151 & 0.01416 & 0.11116 & -0.00369 & -- & 0.09819 \\
 \hline
 14 &0.17949 & -0.06135 & 0.01410 & 0.11124 & -0.00366 & -- & 0.09825\\
 \hline \hline
 $\infty$ & 0.178 & -0.061 & 0.014 & 0.112 & -0.004 & -- & 0.098 \\
 \hline
\end{tabularx}
\caption{
Same as Table~\ref{tab:AnglesXCircmp5gp3} except with $m=0.1$ and $g=0.8$
and through 6 steps of the SC-ADAPT-VQE algorithm.}
 \label{tab:AnglesXCircmp1gp8}
\end{table}
\begin{table}[!ht]
\renewcommand{\arraystretch}{1.4}
\begin{tabularx}{\textwidth}{|c || Y  | Y || Y |  Y || Y || c |}
 \hline
 $L$ 
 & $\varepsilon^{\rm (aVQE)}$ 
 & $\varepsilon^{\rm (exact)}$ 
 & $\chi^{\rm (aVQE)}$ 
 & $\chi^{\rm (exact)}$ 
 & $\infiL$  & \# CNOTS/\text{qubit}
  \\
 \hline\hline
 6 & -0.41488& -0.41614 & 0.51372 & 0.52154 & 0.00072 & 29.5\\
 \hline
7 & -0.41869& -0.41996 & 0.51579 & 0.52348 & 0.00071 & 32.1 \\
\hline
8 & -0.42156& -0.42283 & 0.51736 & 0.52495 & 0.00071 & 33.9\\
\hline
9 & -0.42379& -0.42506 & 0.51859 & 0.52609 & 0.00071 & 35.2\\
\hline
10 & -0.42557& -0.42685 & 0.51958 & 0.52701 & 0.00071 & 36.3\\
\hline
11 & -0.42669& -0.42831 & 0.51945 & 0.52776 & 0.00129 & 38.9\\
\hline
12 & -0.42799& -0.42953 & 0.52047 & 0.52838 & 0.00121 & 39.7\\
\hline
13 & -0.42909& -0.43056 & 0.52134 & 0.52891 & 0.00115 & 40.3\\
\hline
14 & -0.43003& -0.43144 & 0.52209 & 0.52937 & 0.00109 & 40.9\\
\hline
\end{tabularx}
\caption{
Same as Table~\ref{tab:Echifmp5gp3} except with 
$m=0.1$ and $g=0.8$ and through 6 steps of the SC-ADAPT-VQE algorithm.
}
 \label{tab:Echifmp1gp8}
\end{table}
\begin{table}[!ht]
\renewcommand{\arraystretch}{1.4}
\begin{tabularx}{\textwidth}{|c || Y | Y | Y | Y | Y ||  Y| Y | Y | Y ||}
 \hline
 \diagbox[height=23pt]{$L$}{$\theta_i$} & $\hat O_{mh}^{V}(1)$
 & $\hat O_{mh}^{V}(3)$ & $\hat O_{mh}^{V}(5)$ & $\hat O_{mh}^{V}(1)$ & $\hat O_{mh}^{V}(7)$ &  $\hat O_{mh}^{S}(0,1)$ & 
 $\hat O_{mh}^{V}(1)$ & $\hat O_{mh}^{S}(0,1)$ & $\hat O_{mh}^{V}(9)$\\
 \hline\hline
 6 & 0.17222 & -0.06236 & 0.01456 & 0.11495 & -0.00409 & 0.07017 & 0.09561 & -- & --\\
 \hline
 7 & 0.17278 & -0.06184 & 0.01433 & 0.11431 & -0.00389 & 0.06947 & 0.09620 & -- & --\\
 \hline
 8 & 0.17316 & -0.06147 & 0.01417 & 0.11388 & -0.00378 & 0.06900 & 0.09659 & -- & -- \\
 \hline
 9 & 0.17344 & -0.06121 & 0.01407 & 0.11357 & -0.00371 & 0.06866 & 0.09688 & -- & --\\
 \hline
 10 & 0.17365 & -0.06101 & 0.01399 & 0.11333 & -0.00366 & 0.06840 & 0.09710 & -- & --\\
 \hline
 11 &0.17300 & -0.06058 & 0.01385 & 0.11216 & -0.00359 & -- & 0.09885 & 0.07047 & --\\
 \hline
 12 & 0.17321 & -0.06048 & 0.01381 & 0.11210 & -0.00357 & -- & 0.09883 & 0.07028 & -- \\
 \hline
 13 & 0.17338 & -0.06039 & 0.01378 & 0.11205 & -0.00355 & -- & 0.09883 & 0.07012 & --\\
 \hline
 14 & 0.17950 & -0.06139 & 0.01417 & 0.11124 & -0.00382 & -- & 0.09825 & -- & 0.00107\\
 \hline 
\end{tabularx}
\caption{
Same as Table~\ref{tab:Echifmp5gp3} except with 
$m=0.1$ and $g=0.8$ and through 7 steps of the SC-ADAPT-VQE algorithm.}
 \label{tab:AnglesXCircmp1gp8Step7}
\end{table}
\begin{table}[!ht]
\renewcommand{\arraystretch}{1.4}
\begin{tabularx}{\textwidth}{|c || Y | Y || Y | Y |}
\hline
\multicolumn{5}{|c|}{$m=0.1, g=0.3$} \\
\hline
 \hline
 $L$ 
 & $\varepsilon^{\rm (SC-MPS)}$ 
 & $\varepsilon^{\rm (DMRG)}$ 
 & $\chi^{\rm (SC-MPS)}$  
 & $\chi^{\rm (DMRG)}$ \\
 \hline\hline
 50 &-0.52640 & -0.52797 & 0.70967 & 0.71444 \\
 \hline
 100 & -0.52838 & -0.52971 & 0.71359 & 0.71504\\
 \hline
 \hline
 \multicolumn{5}{|c|}{$m=0.1, g=0.8$} \\
 \hline\hline
 50 & -0.43886 & -0.43971 & 0.53339 & 0.53361 \\
 \hline
 100 & -0.44058 & -0.44131 & 0.53604 & 0.53444\\
 \hline
200 & -0.44144 & -0.44212 &0.53737 & 0.53485 \\
 \hline
 300 & -0.44173 & -0.442384 &0.53781 &  0.53499 \\
 \hline
 400 & -0.44187 & -0.44252 &0.53803 & 0.53506 \\
 \hline
\end{tabularx}
\caption{
Same as Table~\ref{tab:SQCResults} except with $m=0.1,g=0.3$ and $m=0.1, g=0.8$.}
 \label{tab:SQCResultsApp}
\end{table}

\clearpage
\section{Additional Details and Results From Simulations using IBM's Quantum Computers}
\label{app:qusimDetail}
\begin{figure}[!ht]
    \centering
    \includegraphics[width=0.42\columnwidth]{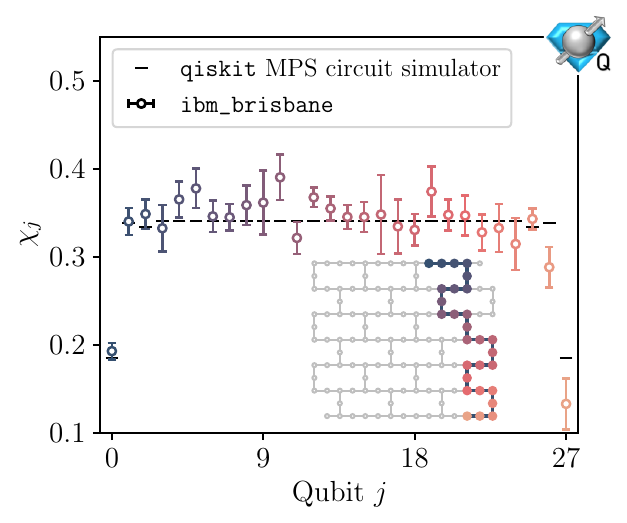}
    \includegraphics[width=0.56\columnwidth]{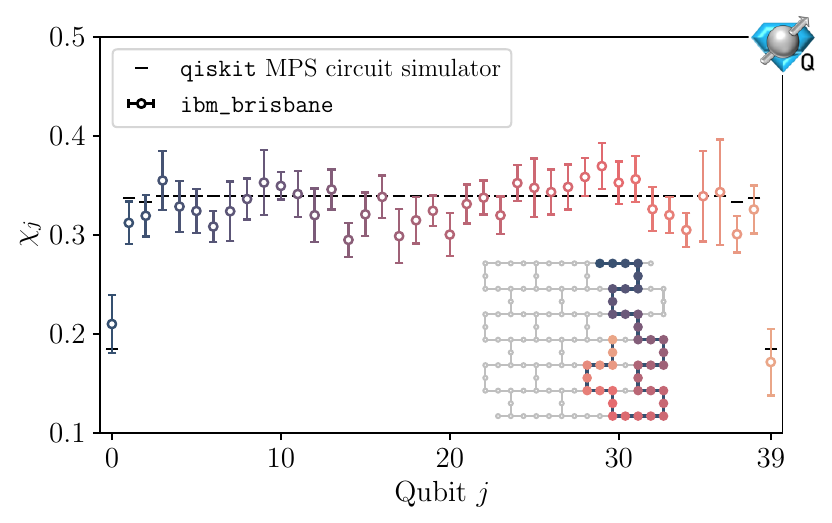}
    \includegraphics[width=0.75\columnwidth]{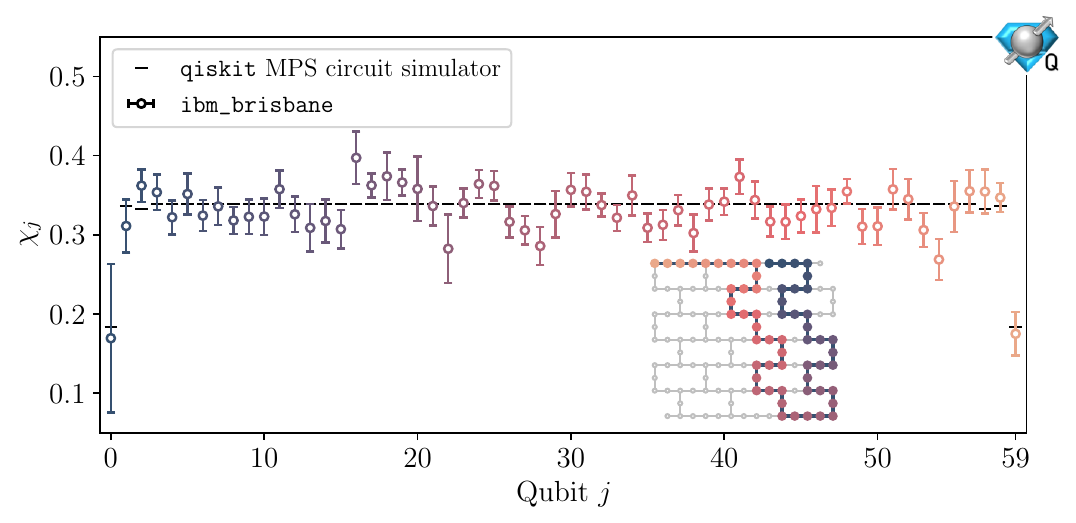}
    \includegraphics[width=0.95\columnwidth]{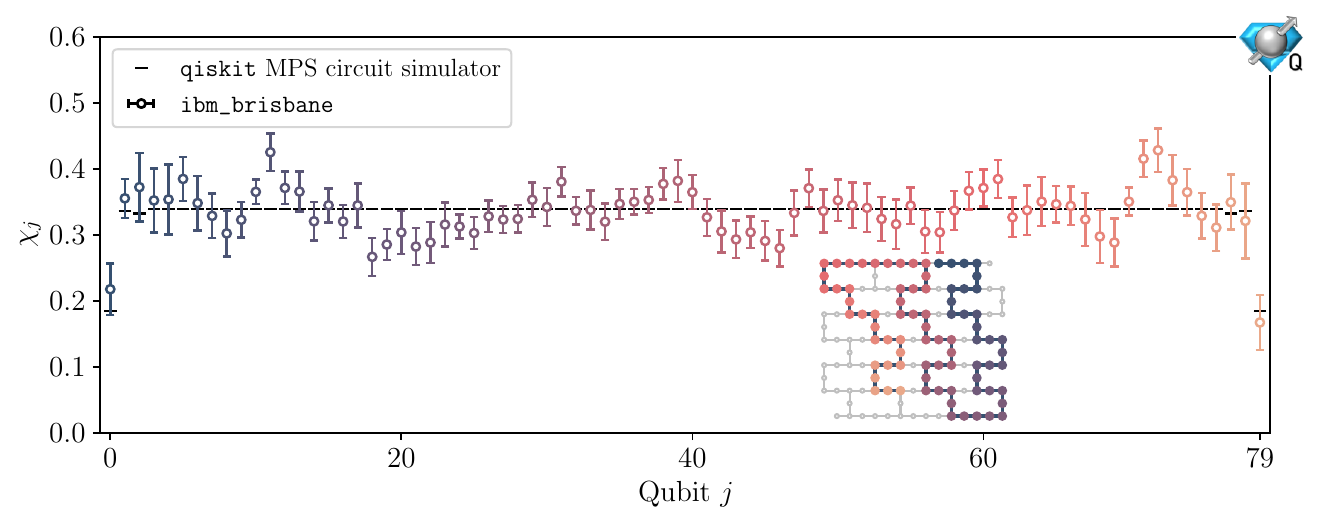}
    \caption{Expectation values of $\chi_j$ for $L=14,20,30$ and $40$ (top to bottom), 
    obtained from simulations using {\tt ibm\_brisbane}. 
    They are compared with the expected results obtained by using {\tt qiskit}'s MPS circuit simulator (black dashes). 
    Averaging $\chi_j$ over all of the 
    qubits provides the chiral condensates presented in Table~\ref{tab:SQCResults_ibm}.
    The layouts of the qubits used on the chip are shown in the insets.}
    \label{fig:chi_14-40}
\end{figure}

\begin{figure}[!ht]
    \centering
    \includegraphics[width=0.37\columnwidth]{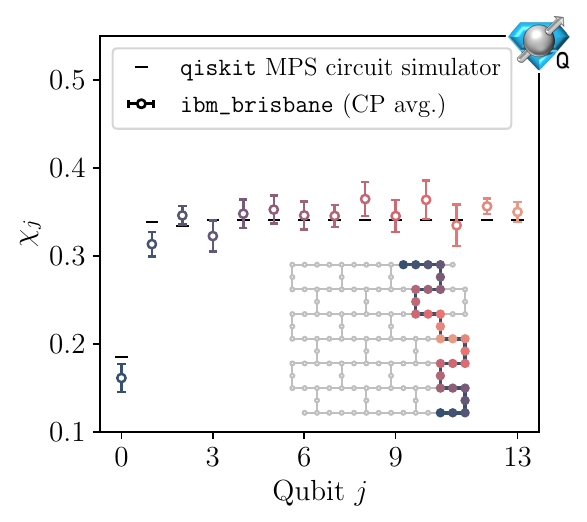}
    \includegraphics[width=0.4\columnwidth]{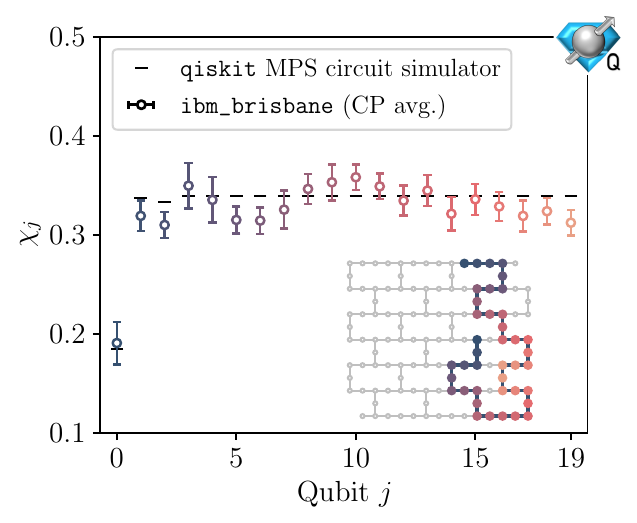}
    \includegraphics[width=0.46\columnwidth]{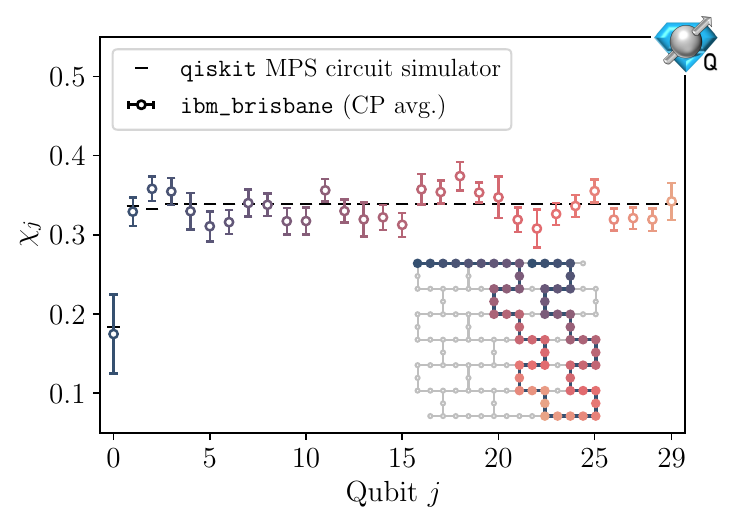}
    \includegraphics[width=0.52\columnwidth]{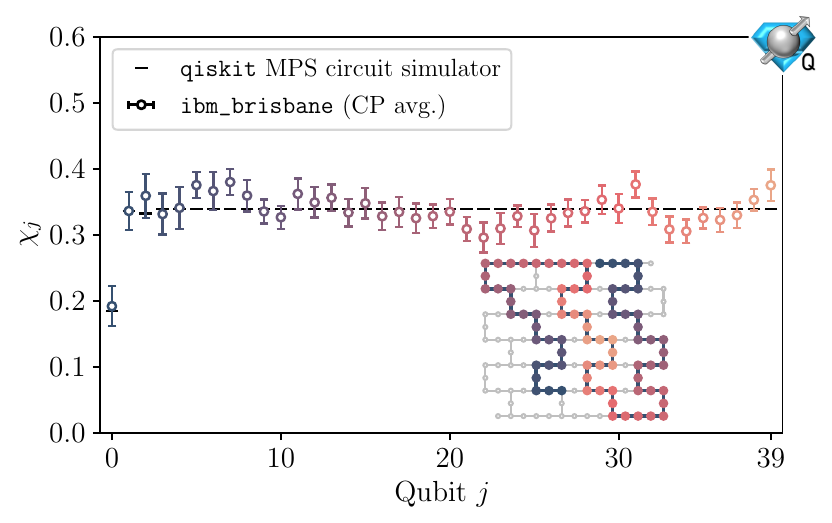}
    \includegraphics[width=0.63\columnwidth]{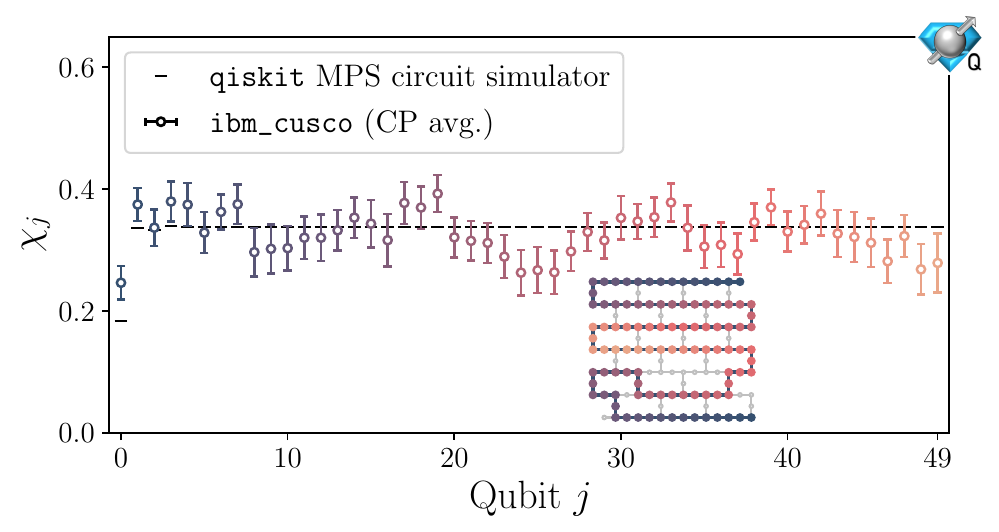}
    \caption{Expectation values of CP averaged $\chi_j$ for $L=14,20,30,40$ and $50$ (top to bottom)
    obtained from simulations using {\tt ibm\_brisbane} and {\tt ibm\_cusco}. 
    They are compared with the expected results obtained by using {\tt qiskit}'s MPS circuit simulator (black dashes). 
    The layouts of the qubits used on the chip are shown in the insets (with same-colored qubits being averaged).}
    \label{fig:chi_14-50_CP}
\end{figure}

\begin{figure}[!ht]
    \centering
    \includegraphics[width=0.92\columnwidth]{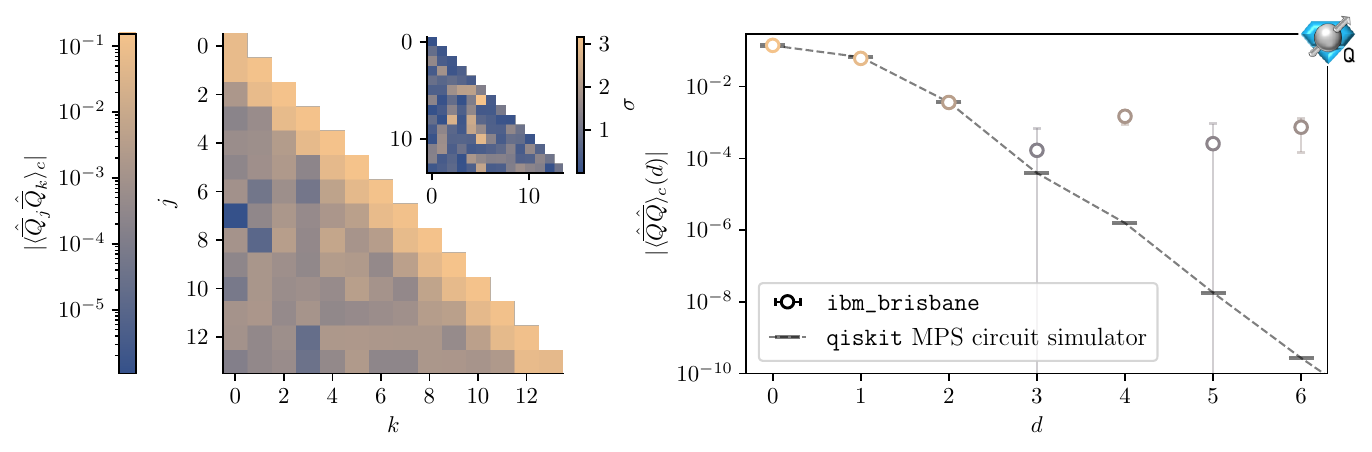}
    \includegraphics[width=0.92\columnwidth]{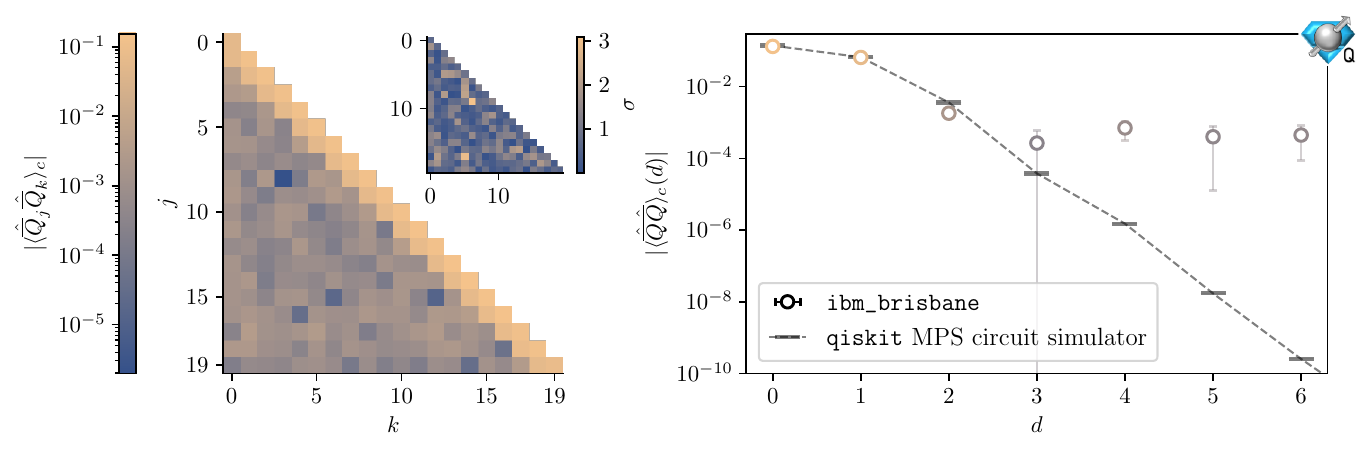}
    \includegraphics[width=0.92\columnwidth]{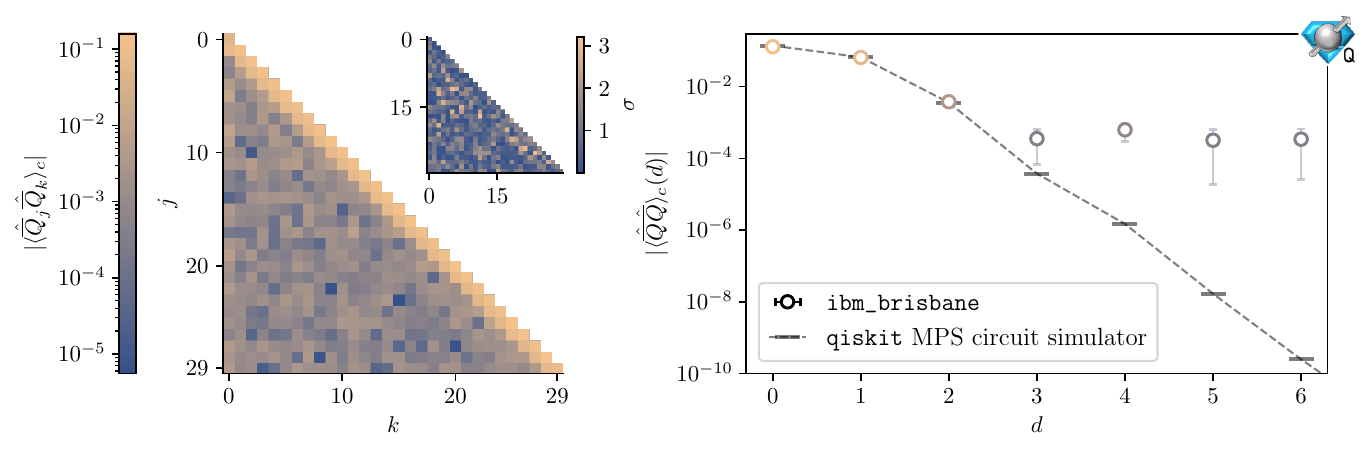}
    \includegraphics[width=0.92\columnwidth]{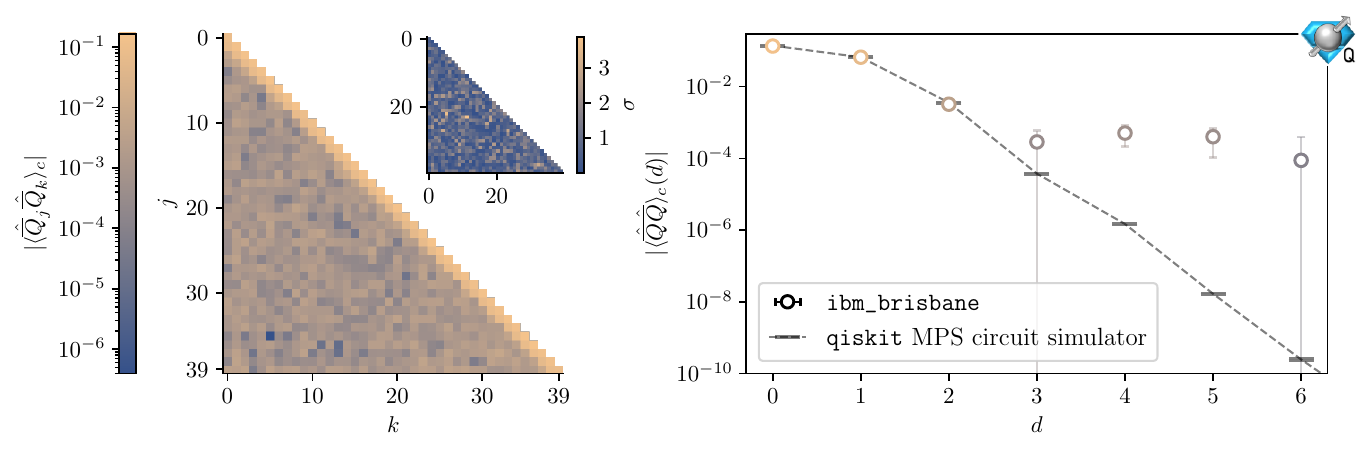}
    \caption{The connected contribution to the spatial charge-charge correlation functions, $\langle \hat{\overline{Q}}_{j} \hat{\overline{Q}}_{k} \rangle_c$ (left)
    and the averaged correlation functions as a function of distance $d$, 
    $\langle \hat{\overline{Q}} \hat{\overline{Q}} \rangle_c (d)$ (right) with the points following the same color map as in the left main panel (error bars show $1\sigma$ standard deviations).
    Results obtained from {\tt ibm\_brisbane} are shown for $L=14,20,30$ and $40$ (top to bottom).}
    \label{fig:qiqj_14-50}
\end{figure}

\noindent
In this appendix, we provide additional details about how our results are obtained from IBM's quantum computers, together with the additional figures not shown in Sec.~\ref{sec:SCQuSim}.
All measurements are performed on {\tt ibm\_brisbane} ($L\leq 40$) and {\tt ibm\_cusco} ($L=50$) by sending the state preparation circuits, 
with measurements in the computational ($z$) basis,
via the {\tt qiskit} Runtime Sampler primitive.
The values of the variational parameters obtained from fitting to the exponential form in Eq.~(\ref{eq:thetaextrap}) for 2 steps of SC-ADAPT-VQE are given in Table~\ref{tab:angles_ibm}. The different qubits used for each lattice size can be seen in the insets in Figs.~\ref{fig:chi_50}
and~\ref{fig:chi_14-40}. 
$\chi_j$, obtained from {\tt ibm\_brisbane} for $L=14,20,30$ and 40, is shown in Fig.~\ref{fig:chi_14-40}, and the charge-charge correlation functions are shown
in Fig.~\ref{fig:qiqj_14-50}.
In Fig.~\ref{fig:chi_14-50_CP}, the CP symmetry relating $\chi_j = \chi_{2L-1-j}$ is used to effectively double the number of shots, resulting in statistical error bars that are smaller by a factor of $\sqrt{2}$.

\begin{figure}[!ht]
    \centering
    \includegraphics[width=0.75\columnwidth]{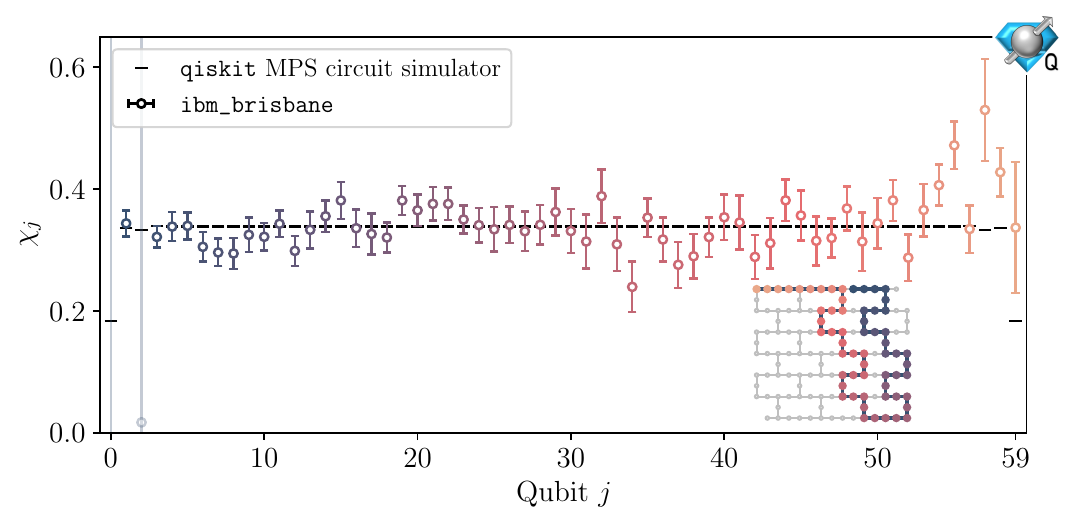}
    \includegraphics[width=0.5\columnwidth]{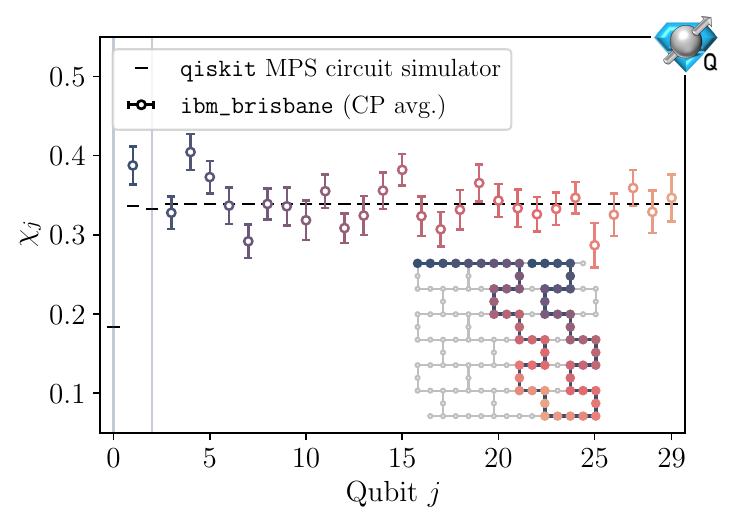}
    \includegraphics[width=0.92\columnwidth]{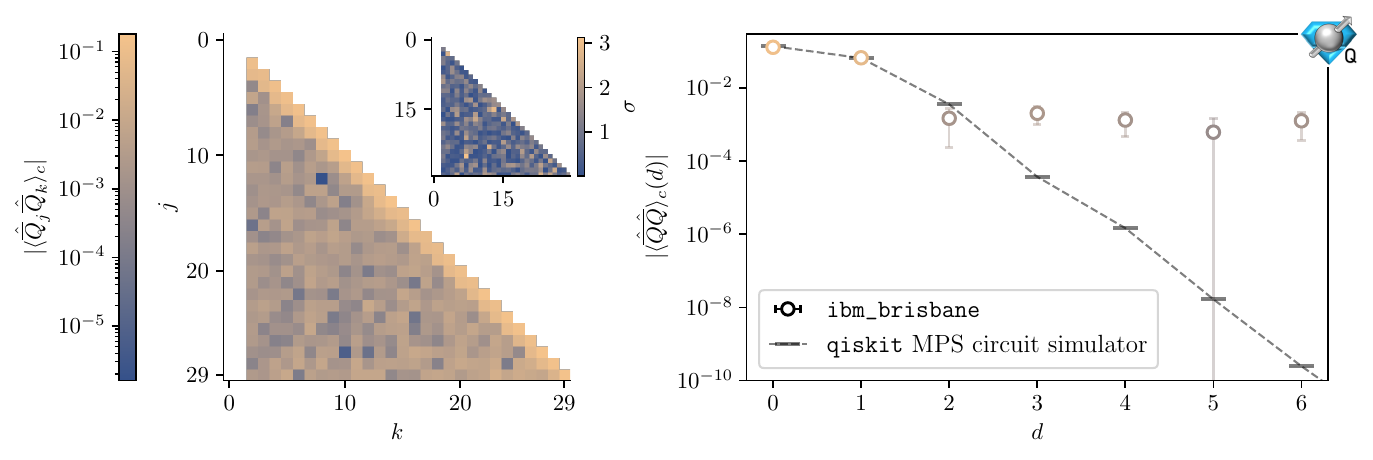}
    \caption{Results for the $L=30$ system obtained with use of three steps of SC-ADAPT-VQE, obtained from simulations using {\tt ibm\_brisbane} with 80 twirled instances. The top panel shows $\chi_j$, the middle panel shows the CP averaged $\chi_j$, and the bottom panels show the connected contribution to the spatial charge-charge correlation functions, $\langle \hat{\overline{Q}}_{j} \hat{\overline{Q}}_{k} \rangle_c$ (the first two spatial sites are not shown due to the errors on qubits 0 and 2),
    and
    the averaged correlation functions as a function of distance $d$, 
    $\langle \hat{\overline{Q}} \hat{\overline{Q}} \rangle_c (d)$, with the points following the same color map as in the left main panel (error bars show $1\sigma$ standard deviations).}
    \label{fig:3layersL30}
\end{figure}

\begin{figure}[!ht]
    \centering
    \includegraphics[width=\columnwidth]{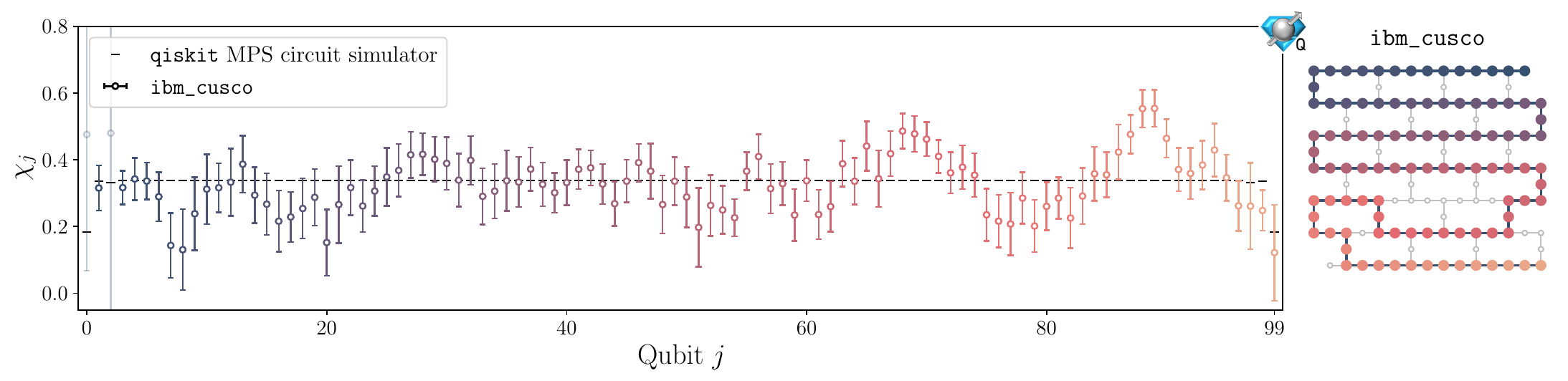}
    \includegraphics[width=0.5\columnwidth]{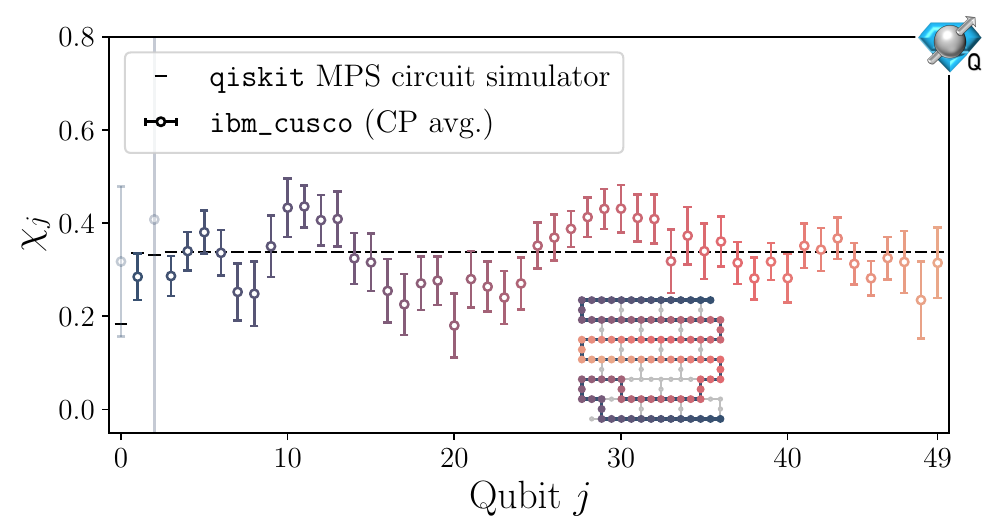}
    \includegraphics[width=0.92\columnwidth]{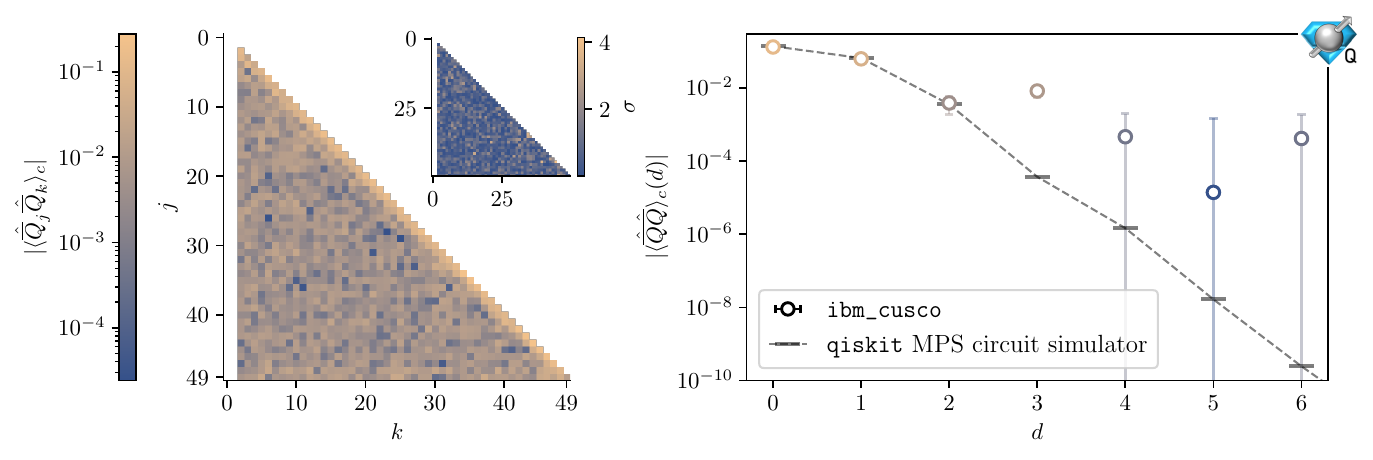}
    \caption{Results for the $L=50$ system obtained with use of three steps of SC-ADAPT-VQE, obtained from simulations using {\tt ibm\_cusco} with 40 twirled instances. The top panel shows $\chi_j$, the middle panel shows the CP averaged $\chi_j$, and the bottom panels show the connected contribution to the spatial charge-charge correlation functions, $\langle \hat{\overline{Q}}_{j} \hat{\overline{Q}}_{k} \rangle_c$ (the first two spatial sites are not shown due to the errors on qubits 0 and 2),
    and
    the averaged correlation functions as a function of distance $d$, 
    $\langle \hat{\overline{Q}} \hat{\overline{Q}} \rangle_c (d)$, with the points following the same color map as in the left main panel (error bars show $1\sigma$ standard deviations).}
    \label{fig:3layersL50}
\end{figure}

In an effort to explore the limitations of the quantum computer, 
the 3-step SC-ADAPT-VQE state preparation circuits for $L=30$ and $L=50$ were implemented on {\tt ibm\_brisbane} and {\tt ibm\_cusco}, respectively. 
The structure of the ansatz wave function and corresponding variational parameters can be found in Table~\ref{tab:angles_ibm}.
The local chiral condensate and charge-charge correlators obtained from 80 ($L=30$) and 40 ($L=50$) twirled instances, each with $8\times 10^3$ shots, are shown in Figs.~\ref{fig:3layersL30} and~\ref{fig:3layersL50}. 
Despite the factor of three increase in the number of
CNOTs relative to 2 layers (1254 versus 468 for $L=30$, and 2134 versus 788 for $L=50$), the results are consistent with those obtained from the {\tt qiskit} MPS circuit simulator.
Note that qubit 0 and 2 have decohered for both volumes, and in principle could be removed
from volume averaged quantities, such as the chiral condensate.

\begin{table}[!ht]
\renewcommand{\arraystretch}{1.4}
\begin{tabularx}{0.6\textwidth}{|c || Y | Y || Y | Y | Y |}
 \hline
 & \multicolumn{2}{c||}{2 steps}  & \multicolumn{3}{c|}{3 steps} \\\hline
 \diagbox[height=23pt]{$L$}{$\theta_i$} 
 & $\hat{O}^{V}_{mh}(1)$ & $\hat{O}^{V}_{mh}(3)$ & $\hat{O}^{V}_{mh}(1)$ & $\hat{O}^{V}_{mh}(3)$ & $\hat{O}^{V}_{mh}(5)$ \\
 \hline\hline
 14 & 0.30699 & -0.04033 & & & \\
 \hline
 20 & 0.30638 & -0.03994 & & & \\
 \hline
 30 & 0.30610 & -0.03978 & 0.30630 & -0.04092 & 0.00671 \\
\hline
 40 & 0.30605 & -0.03975 & & & \\
\hline
 50 & 0.30604 & -0.03975 & 0.30624 & -0.04089 & 0.00670 \\
 \hline
\end{tabularx}
\caption{
Extrapolation of the variational parameters corresponding to 2 and 3 steps of SC-ADAPT-VQE with $m=0.5, g=0.3$.
These parameters were used in the circuits run on {\tt ibm\_brisbane} ($L\leq 40$) and {\tt ibm\_cusco} ($L= 50$). }
 \label{tab:angles_ibm}
\end{table}

By sending the circuits with the Sampler primitive, several error mitigation techniques are applied during runtime, as mentioned in Sec.~\ref{sec:SCQuSim}. Specifically, the readout mitigation technique used (for $L\leq 40$) is M3~\cite{Nation:2021kye}. This method is based on correcting only the subspace of bit-strings observed in the noisy raw counts from the machine (which usually include the ideal ones plus those with short Hamming distance, introduced by the noise in the measurement), and using Krylov subspace methods to avoid having to compute (and store) the full assignment matrix.

Unlike the other works that have utilized $\geq 100$ superconducting qubits~\cite{Yu:2022ivm,Kim:2023bwr,Shtanko:2023tjn}, which used zero-noise extrapolation (ZNE)~\cite{Li:2016vmf,Temme:2016vkz,2020arXiv200510921G} in conjunction with probabilistic error correction (PEC)~\cite{Temme:2016vkz,Berg:2022ugn} to remove incoherent errors, 
we use Operator Decoherence Renormalization (ODR), as explained in App.~\ref{app:QSimError}. 
Both methods require first transforming coherent errors into incoherent errors, which is done via Pauli twirling. However, the overhead in sampling using ZNE and PCE, compared with ODR, is substantial. For ZNE, one has to add two-qubit gates to increase the noise level, and then perform an extrapolation to estimate the noiseless result. In the minimal case, this leads to running only another circuit, like in ODR, but with a circuit depth that is three times as large as the original circuit (e.g., replacing each CNOT with 3 CNOTs). 
However, this leads to a large uncertainty in the functional form of the extrapolation, and ideally the circuit is run with multiple noise levels to have multiple points from which to extrapolate. For PEC, the overhead is even larger, as it involves learning the noise model of the chip, by running multiple random circuits with different depths (see Ref.~\cite{Berg:2022ugn}). For ODR, as explained in App.~\ref{app:QSimError}, only the same ``physics" circuits are run, but with all rotations set to zero, meaning the sampling overhead is only doubled.

To generate the different twirled circuits, the set of two-qubit Pauli gates $G_2$ and $G'_2$ that leave the (noisy) two-qubit gate invariant (up to a global phase) must be identified. For the quantum processors used in this work, the native two-qubit gate is the echoed cross-resonance (ECR) gate, which is equivalent to the CNOT gate via single qubit rotations. Explicitly,
\begin{equation}
ECR = \frac{1}{\sqrt{2}}(\hat{X}\otimes\hat{I}-\hat{Y}\otimes \hat{X})\ , \quad \quad
\begin{quantikz} 
& \gate[wires=2]{ECR} & \qw \\ 
& & \qw
\end{quantikz} =
\begin{quantikz} 
& \qw & \ctrl{1} & \gate{R_z(-\frac{\pi}{2})} & \gate{R_y(\pi)} & \qw \\
& \gate{R_x(\frac{\pi}{2})} & \targ & \qw & \qw & \qw & \qw
\end{quantikz} \ .
\end{equation}
Using the functions from the package {\tt qiskit\_research}~\cite{the_qiskit_research_developers_and_contr_2023_7776174}, together with the two-Pauli gate set shown in Table~\ref{tab:ecrtwirl}, a total of 40 (150) twirled circuits for both mitigation and physics were generated for $L\leq 40$ ($L=50$), each with $8\times 10^3$ shots.
\begin{table}[!t]
\renewcommand{\arraystretch}{1.4}
\begin{tabularx}{\textwidth}{| Y | Y | Y | Y | Y | Y | Y | Y |}
 \hline
 $(\hat{I}\otimes\hat{I},\hat{I}\otimes\hat{I})$ & $(\hat{I}\otimes\hat{X},\hat{I}\otimes\hat{X})$ & $(\hat{I}\otimes\hat{Y},\hat{Z}\otimes\hat{Z})$ & $(\hat{I}\otimes\hat{Z},\hat{Z}\otimes\hat{Y})$ &
 $(\hat{X}\otimes\hat{I},\hat{Y}\otimes\hat{X})$ & $(\hat{X}\otimes\hat{X},\hat{Y}\otimes\hat{I})$ & $(\hat{X}\otimes\hat{Y},\hat{X}\otimes\hat{Y})$ & $(\hat{X}\otimes\hat{Z},\hat{X}\otimes\hat{Z})$ \\
 \hline
 $(\hat{Y}\otimes\hat{I},\hat{X}\otimes\hat{X})$ & $(\hat{Y}\otimes\hat{X},\hat{X}\otimes\hat{I})$ & $(\hat{Y}\otimes\hat{Y},\hat{Y}\otimes\hat{Y})$ & $(\hat{Y}\otimes\hat{Z},\hat{Y}\otimes\hat{Z})$ &
 $(\hat{Z}\otimes\hat{I},\hat{Z}\otimes\hat{I})$ & $(\hat{Z}\otimes\hat{X},\hat{Z}\otimes\hat{X})$ & $(\hat{Z}\otimes\hat{Y},\hat{I}\otimes\hat{Z})$ & $(\hat{Z}\otimes\hat{Z},\hat{I}\otimes\hat{Y})$ \\
 \hline
\end{tabularx}
\caption{Two-Pauli gate set $(G_2,G'_2)$ used to generate the twirled ECR gates, $G'_2 \cdot ECR \cdot G_2 = ECR$.}
 \label{tab:ecrtwirl}
\end{table}

From Fig.~\ref{fig:chi_50}, the effects of each error mitigation method can be seen. 
The first set of results shown are semi-raw, obtained directly from the quantum computer. They are not raw since DD is integrated into the circuits that are run on the machine (REM is also included for $L\leq 40$).
To check the effect that DD has, several runs were performed without it, and a degradation of the signal was visible when qubits were idle for long periods (the effects of not using DD were more evident when the deeper 3-step circuit was run). 
Regarding REM, while the final fully-mitigated results for $L=50$ (no REM applied) and $L \leq 40$ (REM applied) systems are similar in quality, a larger statistical sample for $L=50$ was required to achieve an equivalent level of precision ($2.4\times10^6$ vs $6.4\times 10^5$ shots). 
The second set shows the effects of applying PT (the results for no Pauli twirling corresponded to one twirled instance).
It is seen that all the coherent noise on the different qubits has been transformed into uniform incoherent noise.
The last set shown is after ODR has been used to remove the incoherent noise.

%

\bibliography{bibi}
\end{document}